\documentclass[twocolumn]{aastex631}

\usepackage{amsmath}

\newcommand{\castro}{{\sf Castro}}
\newcommand{\maestroex}{{\sf MAESTROeX}}
\newcommand{\pynucastro}{{\sf pynucastro}}

\newcommand{\amrex}{{\sf AMReX}}
\newcommand{\microphysics}{{\sf Microphysics}}

\newcommand{\kb}{k_B}

\newcommand{\isot}[2]{$^{#2}\mathrm{#1}$}
\newcommand{\isotm}[2]{{}^{#2}\mathrm{#1}}
\newcommand{\proton}{\mathrm{p}}
\newcommand{\neutron}{\mathrm{n}}

\newcommand{\gcc}{\mathrm{g~cm^{-3} }}

\newcommand{\nucleus}{{\tt Nucleus}}
\newcommand{\spinnuclide}{{\tt SpinNuclide}}
\newcommand{\partitionfunction}{{\tt PartitionFunction}}

\newcommand{\composition}{{\tt Composition}}
\newcommand{\rate}{{\tt Rate}}
\newcommand{\reaclibrate}{{\tt ReacLibRate}}
\newcommand{\tabularrate}{{\tt TabularRate}}
\newcommand{\singleset}{{\tt SingleSet}}

\newcommand{\library}{{\tt Library}}
\newcommand{\reacliblibrary}{{\tt ReacLibLibrary}}
\newcommand{\tabularlibrary}{{\tt TabularLibrary}}
\newcommand{\ratecollection}{{\tt RateCollection}}

\newcommand{\amrexastrocxxnetwork}{{\tt AmrexAstroCxxNetwork}}

\newcommand{\derivedrate}{{\tt DerivedRate}}
\newcommand{\approximaterate}{{\tt ApproximateRate}}
\newcommand{\ddt}[1]{{\frac{{d#1}}{dt}}}
\newcommand{\rxn}[1]{{\langle { #1} \rangle}}

\newcommand{\subchapprox}{{\tt subch\_approx}}
\newcommand{\burncell}{{\tt burn\_cell}}

\newcommand{\reaclib}{{\sf REACLIB}}
\newcommand{\starlib}{{\sf STARLIB}}

\usepackage{listings}
\definecolor{lbcolor}{rgb}{0.9,0.9,0.9}                                         \definecolor{cosmiclatte}{rgb}{1.0, 0.97, 0.91}                                    
\newcommand{\supnote}[1]{{(supplemental notebook: {\tt #1})}}

\newcommand{\cxx}{C\nolinebreak\hspace{-.05em}\raisebox{.4ex}{\tiny\bf +}\nolinebreak\hspace{-.10em}\raisebox{.4ex}{\tiny\bf +}}

\shorttitle{\pynucastro}
\shortauthors{Smith Clark et al.}

\begin{document}

\title{\pynucastro: A Python Library for Nuclear Astrophysics}

\correspondingauthor{Alexander Smith Clark}
\email{alexander.smithclark@stonybrook.edu}

\author[0000-0001-5961-1680]{Alexander Smith Clark}
\affiliation{Department of Physics and Astronomy,
Stony Brook University,
Stony Brook, NY 11794-3800, USA}

\author[0000-0003-3603-6868]{Eric T. Johnson}
\affiliation{Department of Physics and Astronomy,
Stony Brook University,
Stony Brook, NY 11794-3800, USA}

\author[0000-0002-2839-107X]{Zhi Chen}
\affiliation{Department of Physics and Astronomy,
Stony Brook University,
Stony Brook, NY 11794-3800, USA}

\author[0000-0001-6191-4285]{Kiran Eiden}
\affiliation{Department of Astronomy,
University of California, Berkeley,
Berkeley, CA 94720-3411, USA}

\author[0000-0003-2300-5165]{Donald E.\ Willcox}
\affiliation{Center for Computational Sciences and Engineering,
Lawrence Berkeley National Lab}

\author[0000-0002-5419-9751]{Brendan Boyd}
\affiliation{Department of Physics and Astronomy,
Stony Brook University,
Stony Brook, NY 11794-3800, USA}

\author[0000-0002-8849-9816]{Lyra Cao}
\affiliation{Department of Astronomy,
Ohio State University,
Columbus, OH 43210-1173, USA}

\author[0000-0002-7815-1496]{Christopher J. DeGrendele}
\affiliation{Department of Applied Mathematics,
University of California, Santa Cruz,
Santa Cruz, CA 95064-1077, USA}

\author[0000-0001-8401-030X]{Michael Zingale}
\affiliation{Department of Physics and Astronomy,
Stony Brook University,
Stony Brook, NY 11794-3800, USA}



\begin{abstract}

We describe \pynucastro\ 2.0, an open source library for interactively creating and exploring astrophysical nuclear reaction networks.  We demonstrate new methods for approximating rates and using detailed balance to create reverse rates, show how to build networks and determine whether they are appropriate for a particular science application, and discuss the changes made to the library over the past few years. Finally, we demonstrate the validity of the networks produced and share how we use \pynucastro\ networks in simulation codes.
\end{abstract}



\section{Introduction} \label{sec:intro}

We present \pynucastro\ 2.0, the latest version of our open source python library for nuclear astrophysics\footnote{\url{https://github.com/pynucastro/pynucastro/}}.  \pynucastro\ began as
a simple python library that could interpret the reaction
rate format used by \reaclib~\citep{reaclib} and output
the python code to integrate a network linked by reaction
rates from that library.  The initial intent was for use in a classroom, to make it easier for students to work with and interactively explore reaction networks.  It quickly grew beyond that role, as we recognized the need to explore reaction rates, determine which reactions are important in different environments and generate networks for multidimensional hydrodynamics simulation codes.  Nuclear experiments are constantly refining our understanding of nuclei and reactions \citep{horizons}, and we want to be able to use the most up-to-date reaction rates in our simulations without having to manually update the simulation codes.
\pynucastro\ meets this need by interfacing with compilations of nuclear reaction rates and nuclear properties generating the python or \cxx\ code needed to integrate a reaction network in a simulation code.  \pynucastro's object oriented design
makes it easy to expand its capabilities by building on existing rate or network classes.  The development process is fully opened, managed on Github through pull-requests and issues.

Since the the 1.0 release of \pynucastro\ \citep{pynucastro}, a lot of new features were added, including \cxx\ code generation, the ability to modify and approximate rates, nuclear partition functions and the derivation of reverse rates via detailed balance, support for weak rate tables, nuclear statistical equilibrium state determination, electron screening support, many new plot types, and Numba acceleration of python networks. Finally, \pynucastro\ can generate the reaction networks needed for the \amrex-Astrophysics suite of simulation codes~\citep{amrex-astro}, and leverage their ability to offload onto GPUs for accelerated computing. In this paper, we describe these new features, show examples of how we use the library, and demonstrate its validity.

The general evolution of a specie involved in a two body reaction rate, $A + B \rightarrow C + D$, takes
the form:
\begin{equation} \label{eq:reaction_react}
    \ddt{n_A} = \ddt{n_B} = - (1 + \delta_{AB}) n_A n_B \frac{\rxn{\sigma v}}{1 + \delta_{AB}} \\
\end{equation}
where $n_A$ and $n_B$ are the number densities of species $A$ and $B$ and $\rxn{\sigma v}$ is the reaction rate, expressed as a average of the product of the cross section, $\sigma$, and the relative velocity, $v$ (assumed to follow a Maxwellian distribution), and the denominator corrects for doubling counting if particles $A$ and $B$ are identical
(see, for example, \citealt{clayton,arnett}).
The minus sign here signifies that species $A$
and $B$ are destroyed in this reaction, and the factor of $(1 + \delta_{AB})$ in the numerator indicates that if $A$ and $B$ are the same, then two of the nucleus are destroyed in the reaction.  A corresponding equation for products $C$ and $D$ will have a plus sign:
\begin{equation}\label{eq:reaction_prod}
    \ddt{n_C} = \ddt{n_D} = + n_A n_B \frac{\rxn{\sigma v}}{1 + \delta_{AB}} 
\end{equation}
In terms of molar fraction, $Y_i \equiv n_i m_u / \rho$, we have, for example for species $A$:
\begin{equation}
    \label{eq:molar_evolve}
    \ddt{Y_A} = - (1 + \delta_{AB}) \rho Y_A Y_B \frac{N_A \rxn{\sigma v}}{1 + \delta_{AB}}
\end{equation}
where $N_A$ is Avogadro's number.  The form $N_A \rxn{\sigma v}$ is the quantity that reaction rate tabulations usually provide (as a fit or a table in terms of temperature, $T$), for example, as in the classic compilation \citet{caughlan-fowler:1988}.  There are other variations to this, including for three-body reactions and decays, but the general idea is the same---the change in species molar fraction is given by an ordinary differential equation (ODE) that depends on the molar fraction of reactants, the reaction rate, and density.  Reaction networks usually work in terms of 
molar fractions, since it is easy to match the coefficients in the ODEs to the stoichiometric values from the reaction balance (see \citealt{hix_meyer} for a review of reaction rates).

A reaction network is the collection of nuclei and the rates that link them together, and is expressed as a system of ODEs, $\{dY_k/dt\}$. Each term on the right-hand side of the ODE represents a link between the k-th nucleus (with mass fraction $Y_k$) and another nucleus in the network.  Astrophysical simulations use networks ranging from a few nuclei to several thousand, and managing and integrating the reaction network can be computationally challenging (see, for example, \citealt{timmes_networks}).  Nuclear timescales can be short, and often the reaction network is integrated using implicit methods, requiring a Jacobian to solve the linearized system representing a single step.  A network can evolve the species alone, at fixed temperature and density, include a temperature/internal energy equation under the assumption of constant density or pressure (a self-heating network, as in, e.g., \citealt{chamulak:2008}), evolve according to a prescribed thermodynamic trajectory (like is often done with tracer particles, e.g., \citealt{ivo_tracers}), be used in an operator-split fashion with a hydrodynamics code, with or without an energy equation integrated together with the species, e.g.,  \citet{muller:1986}, or fully coupled to the hydrodynamics equations and include the velocity-dependent advective terms in addition to energy evolution (like, for example, done in \citealt{castro_simplified_sdc}).

\pynucastro\ is designed to aid in the creation and management of reaction networks.  It knows how to compute the reaction rates, $N_A\langle \sigma v \rangle$, and assemble the terms in (\ref{eq:molar_evolve}) that describe a reaction's contribution to the evolution of a nucleus.  The goal is to produce a network that can be used in a simulation with just a few lines of python, with the library pulling in the up-to-date reaction rates, finding all of the links, and writing out the code that represents the righthand side of the system of ODEs.
Additionally, since we usually want to use a minimally-sized network for multi-dimensional simulations, \pynucastro\ understands rate approximations and
has methods for determining if specific rates are important for a particular application.

There are already a number of freely available nuclear reaction networks, including the set of approximate ({\tt aprox13} and {\tt aprox19}) and general ({\tt torch}) networks \citep{timmes_networks}, BRUSLIB/NETGEN \citep{bruslib1,bruslib2} and SkyNet \citep{skynet}.  \pynucastro\ has different goals than these, and as such, provides complementary benefits to the community.  \pynucastro\ encourages interactive exploration of rates and networks in Jupyter notebooks, and is designed to output fixed-sized reaction networks (potentially with approximate rates) that can be used in simulation codes.  Being written completely in python, it has a low barrier of entry for new contributions, and encourages the addition of new features through an open development model.  By writing out the code explicitly for a particular network, and not for a general net that can be configured at runtime, we are able to produce networks that can run efficiently on GPUs as part of large simulation codes \citep{katz:2020}.  Additionally, when coupling to hydrodynamics, we need to integrate the network
with an energy equation and possibly advective terms, which requires more flexibility in how the integration is done than other libraries permit.

This paper is organized as follows.  In section \ref{sec:structure} we describe the overall design of the library.  In section \ref{sec:design_net} we walk through the process of creating a reaction
network and integrating it.  We describe the ability to approximate rates in section \ref{sec:approximate}, discuss tabular rates in section \ref{sec:tabular}, and show how to compute inverse rates via detailed balance in section \ref{sec:actual_detailed_balance}. In sections \ref{sec:screen} and \ref{sec:nse} we describe screening and establishing nuclear statistical equilibrium.  Finally,
in section \ref{sec:comparison} we compare a python and \cxx\ generated network.
A set of Jupyter notebooks is provided as supplementary material archived on zenodo\footnote{\url{https://doi.org/10.5281/zenodo.7202413}} that reproduces all of the figures shown in the paper.  We'll refer to individual notebooks in the figure captions. 

\section{Library Structure}
\label{sec:structure}

\pynucastro\ uses an extensible, object-oriented design and is written completely in python. It leverages Numba just-in-time compilation for acceleration of performance-critical parts, SymPy for \cxx\ code generation, SciPy for integrating python networks, and NetworkX and matplotlib for plotting. We use pytest for unit testing, and have extensive documentation prepared with Sphinx and Jupyter notebooks.  \pynucastro\ is available on PyPI and can easily be installed via {\tt pip}

The library is built from a few core classes that represent nuclei, rates, and
other concepts.  A user works directly with objects of these classes to build
a network and explore it.  The core classes in \pynucastro\ are:

\begin{itemize}
    \item \nucleus: a \nucleus\ represents a single nuclear isotope, and stores the nuclear properties, including proton number, atomic weight, and binding energy.  It also holds \spinnuclide\ and \partitionfunction\ objects that can express the nuclear spin and partition function, respectively.  Finally, a \nucleus\ knows how to print the nucleus to a screen / Jupyter notebook in both plain text and MathJax formatting.
    
    \item \composition: a \composition\ object stores the mass fractions of a collection of nuclei, with methods that enable us to initialize them in various
    fashions (e.g.\ solar), and convert them to molar fractions.  A \composition\ object is needed when evaluating a reaction rate in python.
    
    \item \rate: a \rate\ is a base class that implements functions common to all rate types.  Two main classes are derived from this: \reaclibrate\ for rates from the \reaclib\ library and \tabularrate\ for rates that are specified as a table of $(\rho Y_e, T)$. These classes can evaluate the rate in python and output the source
    code needed to construct the rate in both \cxx\ and python.
    
    Several other classes described below are constructed from these, including
    \approximaterate\ that groups rates together into a single effective rate approximation and \derivedrate\ which recomputes an inverse rate from a forward rate using detailed balance.
    
    \item \library: a \library\ is a collection of \rate\ objects,
      which can be from a particular source (for example, the \reaclib\ database can be read in as a \library).  A \library\ provides many methods for filtering the rates, allowing one to select based on the nuclei involved.
    
      A reaction network is built by selecting one or more rates from a \library.
      
    \item \ratecollection: a \ratecollection\ is the base class for representing a network---a set of \nucleus\ objects and the list of \rate\ objects that link them together.  A \ratecollection\ has a number of methods that help manage a network.  For example, it is at the \ratecollection\ level that one can approximate the network or recompute reverse rates via detailed balance. \ratecollection\ also has many ways to visualize a network, including
    interactively with Jupyter notebook widgets.
    
    The {\tt PythonNetwork} and {\tt AmrexAstroCxxNetwork}
    classes build off of \ratecollection\ to produce the source code for
    a python or \cxx\ network in the \amrex-Astrophysics suite framework.
\end{itemize}

\subsection{Data Sources}

\pynucastro\ uses the \reaclib\ rate library \citep{reaclib} for most rates.  \reaclib\ aims to be a single source for thermonuclear reaction rates, aggregating different community contributions into a single, uniform format.  This provides fits to the $N_A \langle \sigma v\rangle$ portion of the rate in terms of a standard function of temperature.  \reaclib\ does not provide rates for density-dependent electron capture reactions.  For these, we currently include the tables of \citet{suzuki2016}, and other sources can easily be added.

A \reaclib\ \rate\ is described in terms of sets, each of which has 7 coefficients, making up a fit of the form:
\begin{equation}
    \label{eq:reaclib_fit}
    N_A \langle \sigma v \rangle = \exp{\left (a_0 + \sum_{i=1}^5  a_i T_9^{(2i-5)/3}  + a_6 \log T_9\right )}
\end{equation}
with $T_9 = T/(10^{9}\,\mathrm{K})$. A single rate can consist of multiple sets, representing for instance, resonances and the smooth part of the rate, and these are added together to give the total rate.  A \reaclibrate\ stores a python list of \singleset\ objects, each representing a set that contributes to the total rate.  A \singleset\ knows how to numerically evaluate the set as well as output
the python and \cxx\ code to compute the set as part of a reaction network.
The \reacliblibrary\ class reads in the latest stored version of the entire \reaclib\ rate library, and provides the starting point for filtering
out the rates of interest.  

Electron capture and beta-decay rates are taken from \citet{suzuki2016} and stored as a two-dimensional table (electron density and temperature) as part of the \tabularrate\ class.  Linear interpolation is used
to return the needed data at an arbitrary thermodynamic point.
The \tabularlibrary\ class reads in all of the tabular
weak rates known to \pynucastro\ as a library.  It is 
important to note that there can be overlap with tabular
weak rates and approximate weak rates in \reaclib.  Usually
in these cases, one should use the tabular version of the rate (since it captures the density dependence).

Several nuclear properties are needed to derive inverse rates.
In our implementation we use the ground state spin and the mass excesses from \citet{huang:2021} and \citet{wang:2021} under strong experimental arguments. Each rate's $Q$-value is obtained from \reaclib; however, we have implemented an alternative option to compute it directly from the measured mass excesses. The spin measurements may be different from the NuDat database used in \citet{reaclib}, and therefore, we may expect significant differences in heavy or neutron poor/rich nuclei, where systematic measurements dominate. The partition function values are tabulated in \citet{rauscher:1997,rauscher:2003} and merged in the case where each nuclei is present in both tables. We linearly interpolate the logarithm of the partition function from the table data.

\subsection{Development Model}

\pynucastro\ is openly developed on github, with all the changes
done via issues and pull requests. Issues includes new feature requests, bug reports, and questions. Pull requests require a review from a developer along with all tests and workflows passing before they can be merged. Presently, there are over 400 issues + pull requests filed. Developers who have contributed regularly in the past are granted permission to approve and merge pull requests.  Contributors who have made significant contributions were invited to be coauthors of this paper and also appear in the Zenodo record created for each new library release.  Code checkers ({\sf pylint} and {\sf flake8}) are run
on every pull request, as well as a suite of unit tests
(managed through {\sf pytest}), a test compilation and run
of a \cxx\ network with the \amrex-Astro \microphysics\ framework, comparing to
a previously stored answer, and a build of the docs to ensure that changes don't break existing notebook examples.  This is all managed via github actions.  The unit tests currently cover 70\% of the code, with plotting and Numba-wrapped routines as the main exceptions.

\section{Designing a Network in Python}
\label{sec:design_net}

\subsection{Selecting nuclei and rates}

Creating an astrophysical nuclear reaction network is an art form that requires understanding which nuclei are important for your application, and which rates need to be included to accurately capture the energy generation rate.  In \pynucastro\ this translates into assembling the rates that are important into one or more
\library\ objects and then creating a \ratecollection\ (or one of its derived classes).  We often begin
by using a network that is described elsewhere in the literature as inspiration.  As a result of the complexity of ensuring that a network accurately captures all of the processes important to your application, a few networks (like the venerable {\tt aprox13}) dominate most of the multi-dimensional simulation work done today.
\pynucastro\ has tools to help understand what rates and nuclei may be important.

There are a number of ways to select rates for your network:
\begin{itemize}
    \item pass in a list of just the individual rate files you want to use (for example, downloaded from the \reaclib\ website)
    \item read in a large library (like the entire \reaclib\ database) and filter the rates based on which nuclei are involved
\end{itemize}
This will result in a \library\ containing just the rates you want to use.  
But now there are two potential worries:
\begin{itemize}
    \item Some key nuclei or rates might be missing.
    \item The network may be larger than need be if it includes unimportant nuclei and/or rates.
\end{itemize}
A \ratecollection\ provides methods to help with both of these issues.

The easiest way to assess a network is to simply plot it and look to make sure that all of the connections you expect are present.
The {\tt validate()} also helps here---it takes a larger \library\ as an
argument
and it checks the rates in the \ratecollection\ against all of those present
in the comparison library and it will report if any products 
are not consumed by other rates (signifying an endpoint in the network) and if
you are missing any other rates with the exact same reactants (because, for instance, you included protons but not neutrons in your initial rate filtering). An example
of the latter is that you might have $\isotm{C}{12}(\isotm{C}{12}, \proton)\isotm{Na}{23}$ but not $\isotm{C}{12}(\isotm{C}{12}, \neutron)\isotm{Mg}{23}$, because you didn't include neutrons and \isot{Mg}{23} is in the initial list of nuclei you wanted to consider. 

The {\tt find\_unimportant\_rates()} method can help
trim down a network by computing the relative magnitude of each rate in a network. Given a list of thermodynamic states (as a tuple: {\tt (density, temperature, Composition)}) and a threshold, $\epsilon$, it returns a list of all the rates that are always smaller than $\epsilon \times |\mbox{fastest rate}|$.  This can then be used to automatically remove rates.

To illustrate the process of creating a network, we'll show how to make a simple CNO network and discuss the
above considerations.
In the examples that follow we import the \pynucastro\ library as:
\begin{lstlisting}
import pynucastro as pyna
\end{lstlisting}
We will start with the \reaclib\ library.  The class \reacliblibrary\ (derived from \library) reads the entire
library (currently $> 80,000$ rates).
\begin{lstlisting}
rl = pyna.ReacLibLibrary()
\end{lstlisting}
There are a number
of ways to filter rates from the library.  For instance,
if we want just the $\isotm{C}{12}(\proton,\gamma)\isotm{N}{13}$ rate, then we
can do:
\begin{lstlisting}
c12pg = rl.find_rate_by_name("c12(p,g)n13")
\end{lstlisting}
We can then interact with this \rate\ object directly, for example, to plot it or evaluate it at some temperature.

We'll take a different approach for our CNO network.  We can start
with a list of nuclei we are interested in, and then use
{\tt linking\_nuclei()} to select only rates that involve these input nuclei.
\begin{lstlisting}
lib = rl.linking_nuclei(["p", "he4",
                         "c12", "c13",
                         "n13", "n14", "n15",
                         "o15"])
pynet = pyna.PythonNetwork(libraries=[lib])
\end{lstlisting}

\begin{figure}[t]
\plotone{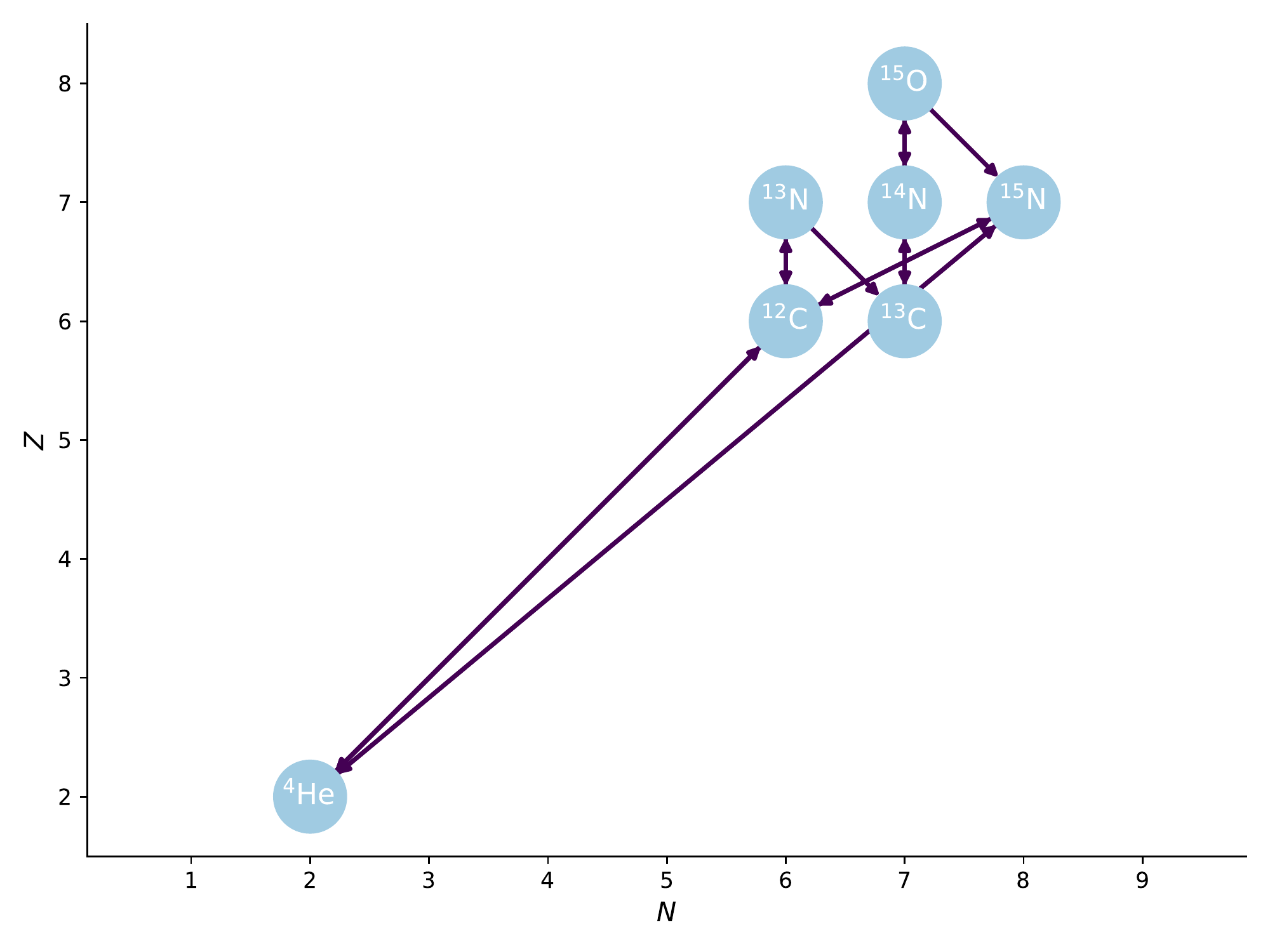}
\caption{\label{fig:cno_first} A simple network made by finding all rates that connect the input nuclei. \supnote{cno-network.ipynb}}
\end{figure}

Optional arguments to {\tt linking\_nuclei} allow one to consider just the reactants or products.  More sophisticated filtering can also be done.
Figure~\ref{fig:cno_first} shows a plot of the network created in this manner (created simply via {\tt pynet.plot()}).  The arrows indicate the direction of the rate, with double-ended arrows meaning both the forward and reverse rates are present (the two can be optionally be plotted as arcs so they don't overlap).
We see that this network
has more rates than we really needed if we just want to consider H burning via CNO.  In conditions where CNO is important, the triple-alpha rate probably is not.  Also, it is unlikely that we will need the reverse rates for the CNO sequence ($\proton+\proton$ is probably important, but we omit that here to simplify the example).

\begin{figure}[t]
\plotone{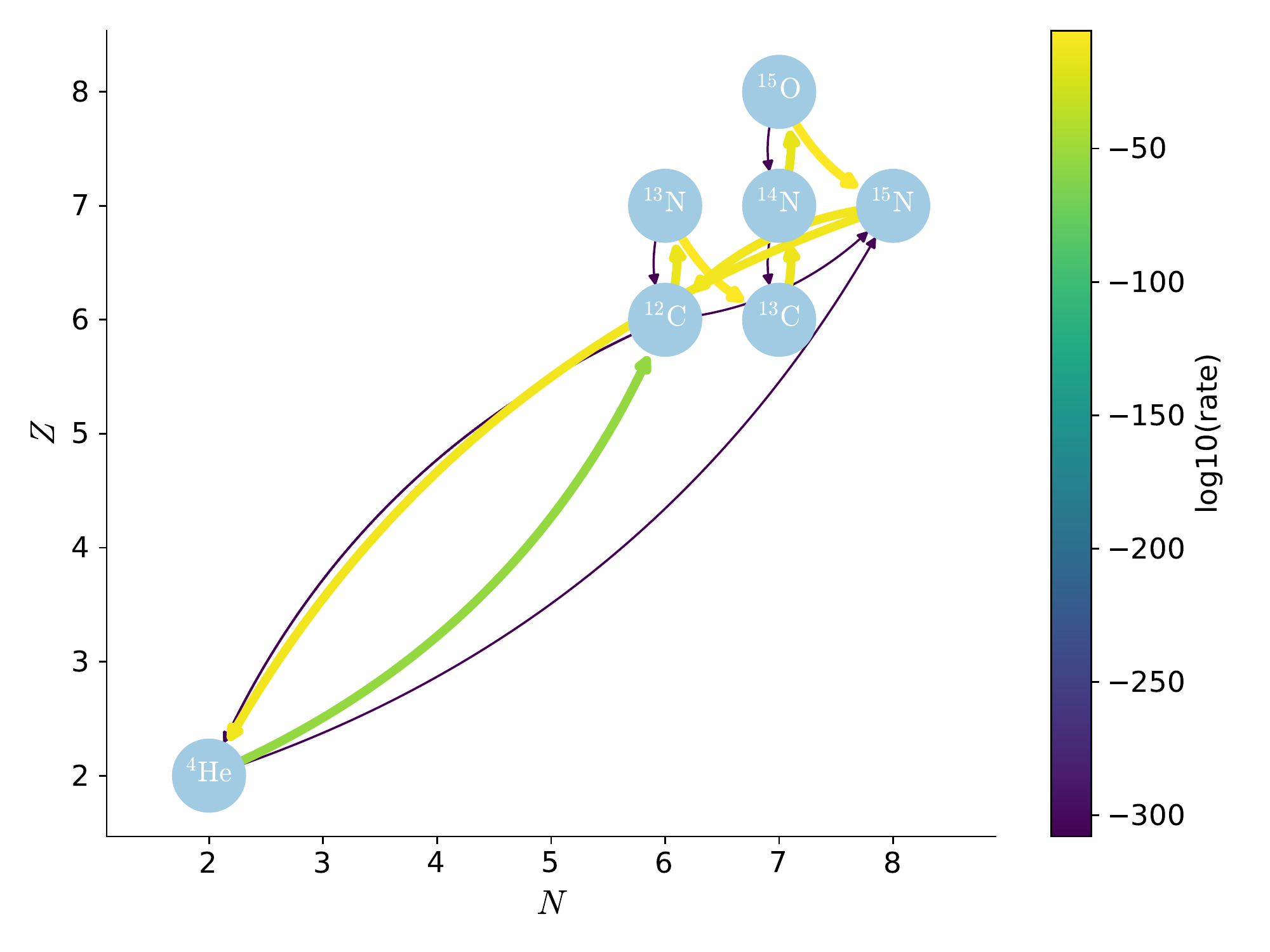}
\caption{\label{fig:cno_colorized} The CNO network with the links colored by reaction rate strength. \supnote{cno-network.ipynb}}
\end{figure}

To simplify this network, we evaluate the rates at a thermodynamic state slightly hotter than the Sun's core temperature and remove the small rates.  We need to build a \composition\ object that stores the mass fractions for this network---we'll assume a solar-like distribution.  First we can look at a plot with the links colored by the reaction rate strength for this thermodynamic state---this is shown in Figure~\ref{fig:cno_colorized}.  As we suspected, the reverse rates are all small, as is the 3-$\alpha$ rate, compared to the main CNO rates.  In fact, on this scale, the rates span 300 orders of magnitude.

\begin{figure}[t]
\plotone{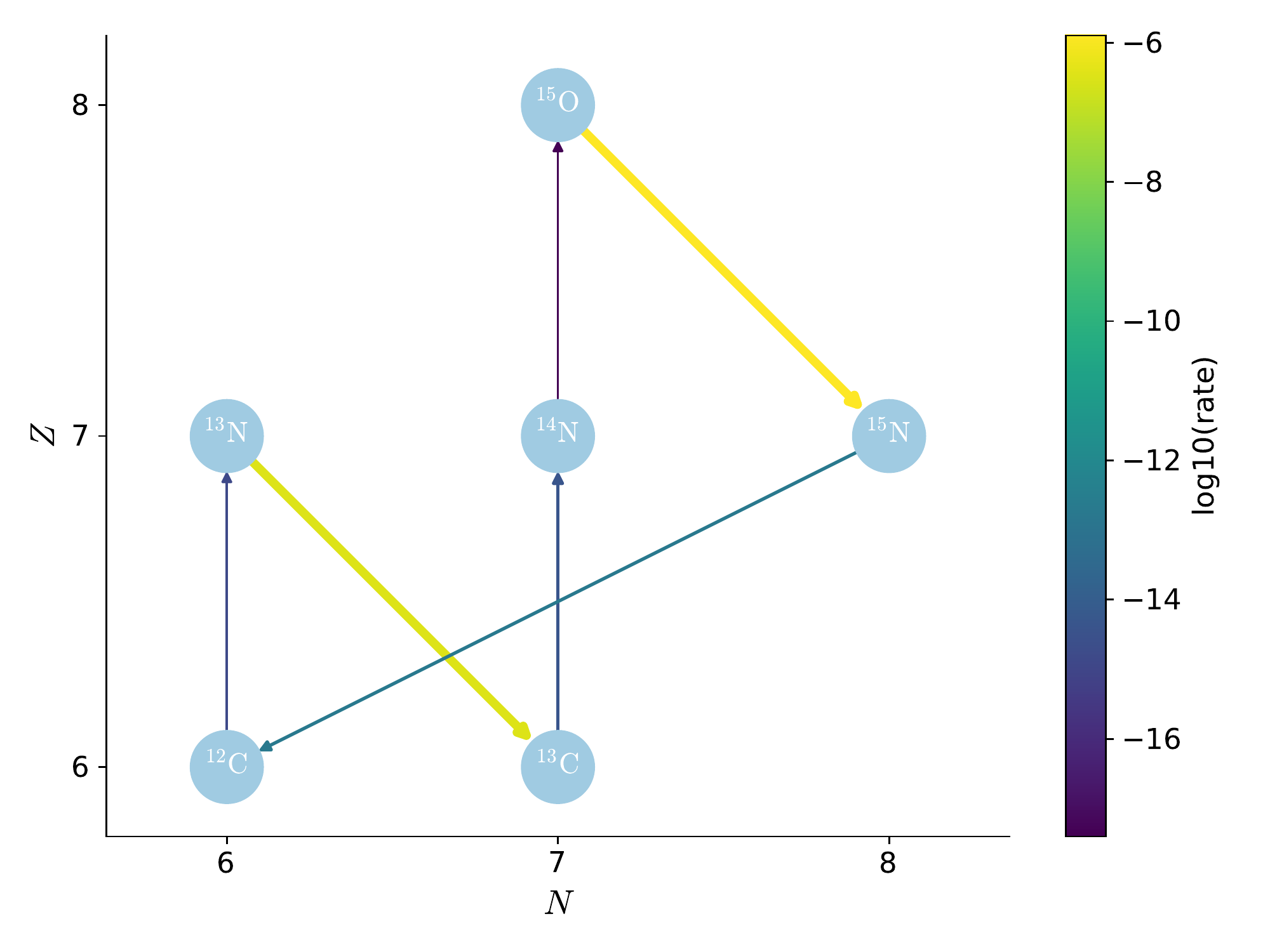}
\caption{\label{fig:cno_filtered} Filtering the \ratecollection\ from Figure~\ref{fig:cno_first} by removing unimportant rates. \supnote{cno-network.ipynb}}
\end{figure}

We can now remove the slow rates.  We'll use an automated method, and filter the rates that are less than $10^{-20}$ compared to the fastest rate.  We don't worry about screening under these conditions, but this check can be done for screened rates as well.
\begin{lstlisting}
comp = pyna.Composition(pynet.get_nuclei())
comp.set_solar_like()
density = 150         # gram/(cm^3)
temperature = 2.e7    # Kelvin
state = (density, temperature, comp)
srates = pynet.find_unimportant_rates([state],
                                      1.e-20)
pynet.remove_rates(srates)
\end{lstlisting}
Figure~\ref{fig:cno_filtered} shows the result.  We note that \isot{He}{4} and $\proton$ are not shown by default, unless $\proton + \proton$ or 3-$\alpha$ are present in the network.  We see that now the range of rates is much more reasonable.  The reverse rates are gone, as is the 3-$\alpha$ rate, and we see
the CNO cycle is all that remains.  It is also clear from the plot that the $\isotm{N}{14}(\proton,\gamma)\isotm{O}{15}$ rate is the limiting rate for the cycle.

This was a simple example, and we could have just specified the exact rates that make up the CNO cycle from the start.  But for larger networks and applications to new problem domains, we might not know what rates are needed.  Furthermore, {\tt find\_unimportant\_rates()} can take a list of thermodynamic states, for instance, sampled from a real simulation at different times and positions in the domain, and remove only the rates that are never important under any of those conditions.

\subsection{Integrating}
\label{sec:integrating}

Once we are happy with the network, we can write out the python code that implements all of the functions needed to integrate the network via {\tt pynet.write\_network()}.  In particular, this will define a module that
contains a function {\tt rhs()} that fills the righthand side of the vector of ODEs
that describe the network ($\dot{Y}_k$), a function {\tt jac()} that creates a Jacobian array ($J_{km} = d{\dot{Y}_k}/dY_m)$, as well as the nuclear properties like proton number, $Z_k$, and atomic mass, $A_k$.   Reaction networks are often stiff, requiring implicit methods, and hence a Jacobian.
Each rate also knows how to write the function that computes its temperature-dependent $N_A\langle \sigma v\rangle$.

\begin{figure}[t]
\centering
\plotone{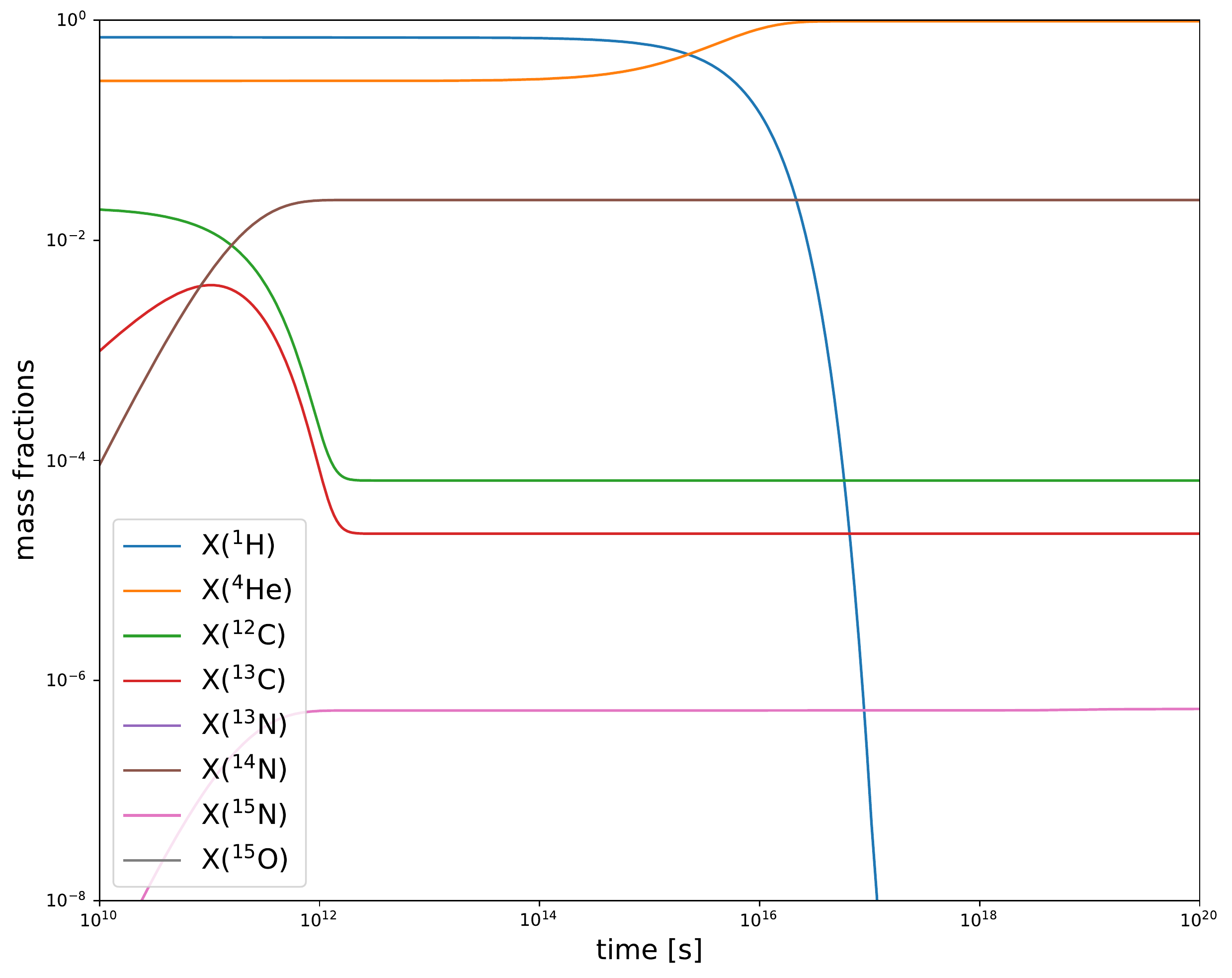}
\caption{\label{fig:cno_integration} Example integration of the CNO network. \supnote{cno-network.ipynb}}
\end{figure}

The following code shows how to write the network
and integrate it using SciPy {\tt solve\_ivp()}.  We create and
initialize an array of mass fractions and then convert to molar fractions, using the nuclear properties in the network module.
This becomes the initial conditions for the integration:
\begin{lstlisting}
pynet.write_network("cno.py")
import cno
rho = 150   # gram/(cm^3)
T = 2.e7    # Kelvins

X0 = np.zeros(cno.nnuc)
X0[cno.jp] = 0.7      # unitless mass fraction
X0[cno.jhe4] = 0.28   # unitless mass fraction
X0[cno.jc12] = 0.02   # unitless mass fraction

Y0 = X0/cno.A

tmax = 1.e20   # seconds

sol = solve_ivp(cno.rhs, [0, tmax], Y0,
                method="BDF",
                jac=cno.jacobian,
                dense_output=True,
                args=(rho, T),
                rtol=1.e-6, atol=1.e-8)
\end{lstlisting}
We use the BDF integration method, since that works well with stiff ODEs and we explicitly
pass in the Jacobian function.  The density and temperature as passed in via the
{\tt args} keyword argument---if we wished, we could easily make these time dependent, computing them on the fly from the current simulation time.  Finally,
we specify both a relative and absolute tolerance, that are combined into a single error of the form: $\epsilon_i = \mathtt{rtol}|X_i| + \mathtt{atol}$.
Figure~\ref{fig:cno_integration} shows the evolution of the nuclei.
For these conditions---slightly hotter than the central temperature
of the Sun---we see that H is depleted in about $10^{17}~\mathrm{s}$.

\section{Approximate Rates}
\label{sec:approximate}

Reaction networks with a lot of nuclei can be computationally expensive---both because of the size of the linear system that
needs to be solved during the implicit integration and when
used in a hydrodynamics code due to the memory needed to carry
and advect each species separately.  As a result, networks are often approximated---removing rates and nuclei deemed not important and replacing a collection of rates with an effective rate.

One of the most common rate approximations is to group the sequence $A(\alpha,\proton)X(\proton,\gamma)B$ and $A(\alpha,\gamma)B$ into a single effective rate.  This is done by assuming that nucleus $X$ and protons are in equilibrium.  The evolution equations
for these two reaction sequences, including forward and reverse rates, is given as:
\begin{subequations}
\begin{align}
   \ddt{Y_A} = \ddt{Y_\alpha} =& 
   - \rho Y_A Y_\alpha \lambda_{\alpha,\gamma}
   + Y_B \lambda_{\gamma,\alpha} \nonumber \\
  &- \rho Y_A Y_\alpha \lambda_{\alpha,\proton}
   + \rho Y_X Y_\proton \lambda_{\proton,\alpha} \\
   \ddt{Y_B} =&
     + \rho Y_A Y_\alpha \lambda_{\alpha,\gamma}
     - Y_B \lambda_{\gamma,\alpha} \nonumber \\
    &+ \rho Y_X Y_\proton \lambda_{\proton,\gamma}
     - Y_B\lambda_{\gamma,\proton} \\
   \ddt{Y_X} = \ddt{Y_\proton} =&
     + \rho Y_A Y_\alpha \lambda_{\alpha,\proton}
     - \rho Y_X Y_\proton \lambda_{\proton,\gamma} \nonumber \\
    &+ Y_B \lambda_{\gamma,\proton}
     - \rho Y_X Y_\proton \lambda_{\proton,\alpha}
\end{align}
\end{subequations}
Here we use the shorthand notation for the rates, with the forward rates:
\begin{subequations}
\begin{align}
    \lambda_{\alpha,\gamma} &= N_A \langle \sigma v \rangle_{A(\alpha,\gamma)B} \\
    \lambda_{\alpha,\proton} &= N_A \langle \sigma v \rangle_{A(\alpha,\proton)X} \\
    \lambda_{\proton,\gamma} &= N_A \langle \sigma v \rangle_{X(\proton,\gamma)B}
\end{align}
\end{subequations}
and reverse rates:
\begin{subequations}
\begin{align}
    \lambda_{\gamma,\alpha} &= N_A \langle \sigma v \rangle_{B(\gamma,\alpha)A} \\
    \lambda_{\gamma,\proton} &= N_A \langle \sigma v \rangle_{B(\gamma,\proton)X} \\
    \lambda_{\proton,\alpha} &= N_A \langle \sigma v \rangle_{X(\proton,\alpha)A}  
\end{align}
\end{subequations}
Finally, notice that the evolution equations for $Y_X$ and $Y_p$ are
identical and the evolution equations for $Y_A$ and $Y_\alpha$
are identical.

If we assume an equilibrium in the proton flow,
$dY_p/dt = 0$, then we can solve for the product $\rho Y_X Y_p$:
\begin{equation}
   \rho Y_X Y_p = \frac{\rho Y_A Y_\alpha \lambda_{\alpha,p} + Y_B \lambda_{\gamma,p}}
                   {\lambda_{p,\gamma} + \lambda_{p,\alpha}}
\end{equation}
Substituting this into the $dY_A/dt$ expression
and grouping the forward and reverse terms together, we have:
\begin{align}
   \ddt{Y_A} = &- \rho Y_A Y_\alpha \left [ \lambda_{\alpha,\gamma} + \lambda_{\alpha,p} \left ( 1 - \frac{\lambda_{p,\alpha}}{\lambda_{p,\gamma} + \lambda_{p,\alpha}} \right ) \right ] \nonumber \\
                   &+ Y_B \left [ \lambda_{\gamma,\alpha} + \frac{\lambda_{p_\alpha} \lambda_{\gamma,p}}{\lambda_{p,\gamma} + \lambda_{p,\alpha}} \right ]
\end{align}
This allows us to identify the effective forward and reverse rates:
\begin{subequations}
\begin{align}
   \lambda^\prime_{\alpha,\gamma} &= \lambda_{\alpha,\gamma} +  \frac{\lambda_{\alpha,p}\lambda_{p,\gamma}}{\lambda_{p,\gamma} + \lambda_{p,\alpha}} \\
   \lambda^\prime_{\gamma,\alpha} &= \lambda_{\gamma,\alpha} +  \frac{\lambda_{p,\alpha}\lambda_{\gamma,p}}{\lambda_{p,\gamma} + \lambda_{p,\alpha}}
\end{align}
\end{subequations}
and the evolution equations reduce to:
\begin{subequations}
\begin{align}
    \ddt{Y_A} = &-\rho Y_A Y_\alpha \lambda^\prime_{\alpha,\gamma} + Y_B \lambda^\prime_{\gamma,\alpha} \\
\ddt{Y_B} = &+\rho Y_A Y_\alpha \lambda^\prime_{\alpha,\gamma} - Y_B \lambda^\prime_{\gamma,\alpha}
\end{align}
\end{subequations}
This approximation is the basis of the popular ``{\tt aprox}'' networks \citep{timmes_networks} and traces its origins at least as far back as \cite{kepler}.  We note that as an effect of this approximation, the protons produced by $(\alpha,p)$ are assumed to only react with the nucleus $X$ and not with any other nuclei in the network.
\pynucastro\ can do this rate approximation automatically.  

\begin{figure*}[t]
\centering
\plottwo{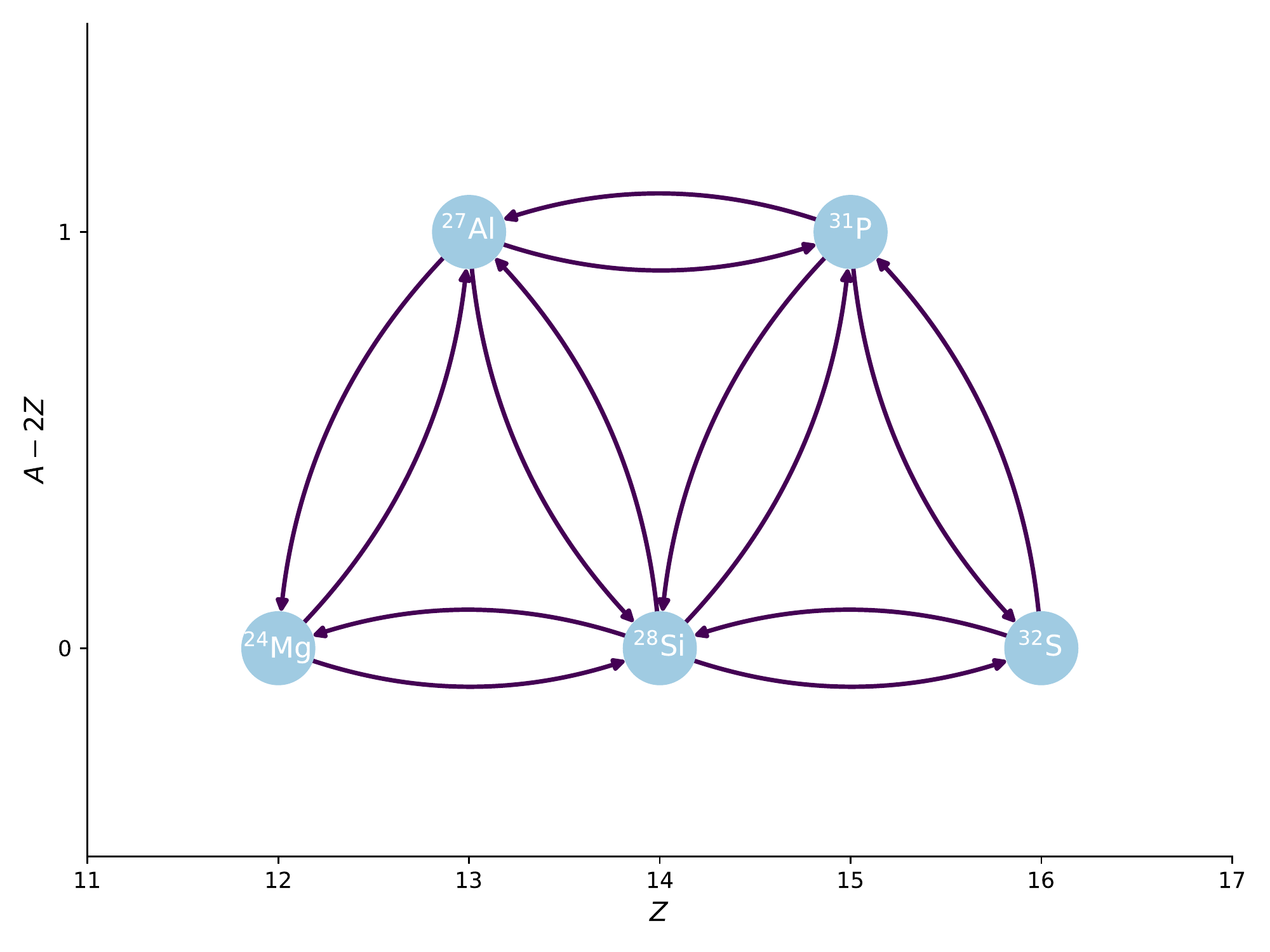}{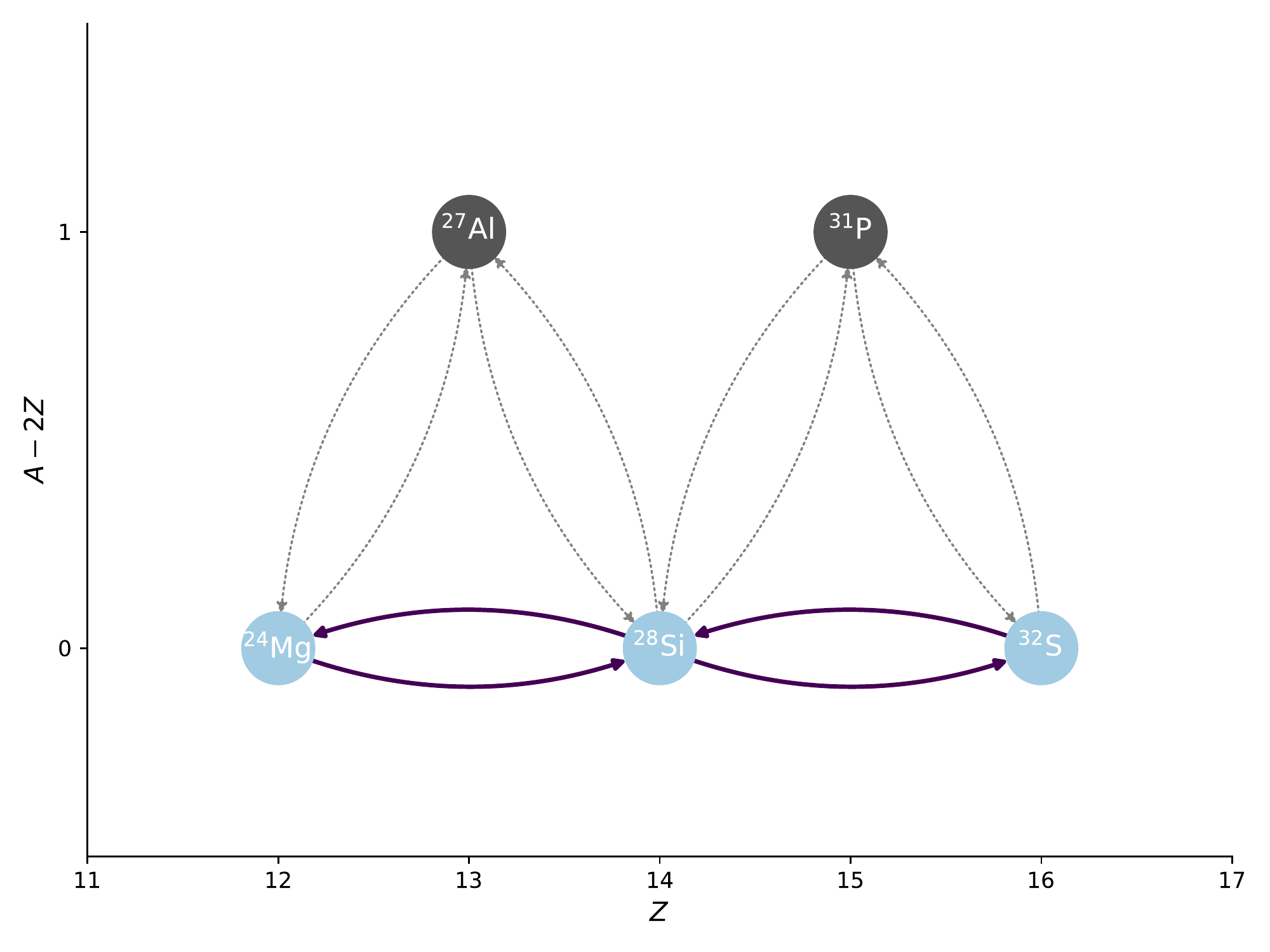}
\caption{\label{fig:approx_diagram} The full \isot{Mg}{24} burning network (left) and the approximate version (right).  Rates that are
approximated are shown as dotted lines and nuclei that are not part of the actual network are shown in gray \supnote{approximate-rates.ipynb}.}
\end{figure*}

Consider a simple network that links \isot{Mg}{24}, \isot{Si}{28}, and \isot{S}{32}
via $(\alpha, p)(p, \gamma)$ and $(\alpha,\gamma)$.  Fully connecting
these nuclei requires that we include \isot{Al}{27} and \isot{P}{31}
as the intermediate nuclei from the $(\alpha,p)$ reactions.  For the approximate version, we'd like to remove those two nuclei from the network, but approximate the flow through them using the effective
rates derived above.  The code to make this approximation is:
\begin{lstlisting}
rl = pyna.ReacLibLibrary()
lib = rl.linking_nuclei(["mg24", "al27",
                          "si28", "p31", "s32",
                          "he4", "p"])
pynet = pyna.PythonNetwork(libraries=[lib])
pynet.make_ap_pg_approx()
pynet.remove_nuclei(["al27", "p31"])
\end{lstlisting}
The method {\tt make\_ap\_pg\_approx()} looks to see if all of the necessary
$(\alpha,\gamma)$, $(\alpha,p)$, and $(p,\gamma)$ rates (and their inverses)
exist in the \ratecollection, and if so, creates an \approximaterate\ object that
replaces the rates with the effective approximation (while storing the original
rates internally to evaluate them for computing the approximation).
We then explicitly remove the intermediate nuclei from the network.
The original
and approximate network is visualized in Figure~\ref{fig:approx_diagram}.  The
gray links in the approximate version indicate that those rates are still
computed and used as part of the approximation.  We note one other feature of
this approximation: there was a link between
\isot{Al}{27} and \isot{P}{31}, but since we removed these nuclei, that link is
also removed.

\begin{figure}[t] 
\centering 
\plotone{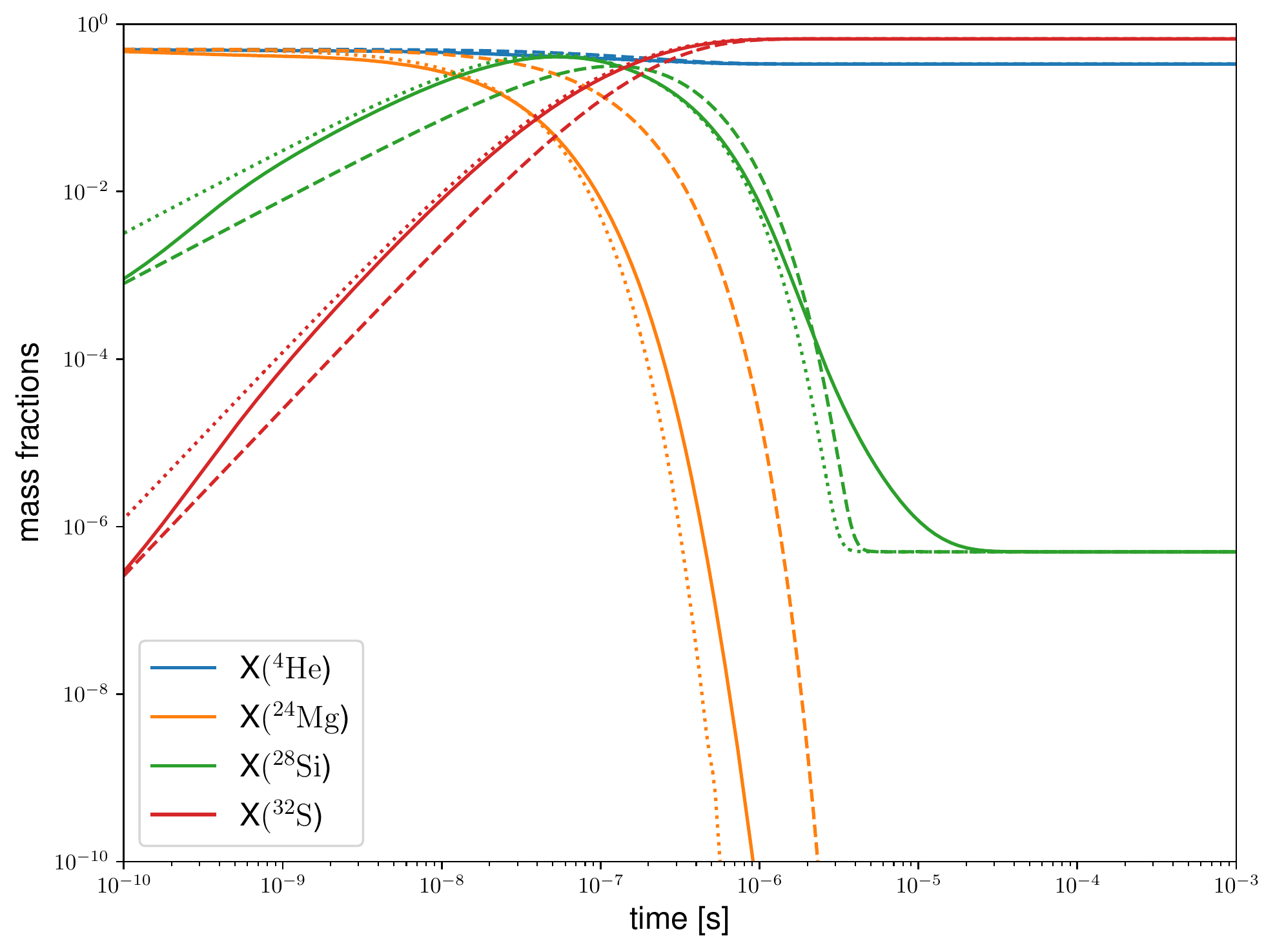}
\caption{\label{fig:approx_results} Comparison of a full network
with $\mathrm{p}$, \isot{He}{4}, \isot{Mg}{24}, \isot{Al}{27}, \isot{Si}{28}, \isot{P}{31}, \isot{S}{32} (solid), an approximate
version that removes \isot{Al}{27} and \isot{P}{31}, by combining
the $(\alpha,p)(p,\gamma)$ and $(\alpha,\gamma)$ rates (dotted) and a
reduced network that has no links (explicit or approximate) to \isot{Al}{27} and \isot{P}{31} (dashed).  Only the 
nuclei in common to all three networks are shown.  We see that the approximate network follows the full network closely, but the reduced network takes longer to burn the initial \isot{Mg}{24} because it is missing the additional pathway through \isot{Al}{27}. \supnote{approximate-rates.ipynb}}
\end{figure}

To see how well this approximation works, we can integrate it along
with the full network (no approximate rates) and a reduced network
with no $(\alpha, p)$ links at all (approximate or explicit).  We pick $\rho = 10^7~\gcc$, $T = 3\times 10^9~\mathrm{K}$, and the initial composition $X(\isotm{He}{4}) = X(\isotm{Mg}{24}) = 0.5$.  The temperature and density are kept fixed.  Figure~\ref{fig:approx_results} shows the comparison.  We see that the network without any $(\alpha,p)(p,\gamma$) links (dashed lines) takes the longest to burn the \isot{Mg}{24} and the trajectories of all of the isotopes is very different from the full network (solid lines).  
In contrast, the approximate network (dotted lines) captures the timescale for \isot{Mg}{24} depletion and \isot{S}{32} creation well, agreeing with the full network after a brief transient.  There is also good agreement in the intermediate \isot{Si}{28} abundance, with a departure only at very low abundances as it is almost depleted.  This shows that the approximate rate infrastructure in \pynucastro\ works as intended.

Approximate networks can be exported both to python and to \cxx.  In this case, \pynucastro\ writes the source code of the functions needed to evaluate each of the rates as well as the code, needed to arrange them together in the approximation, to a file. 
While this is one of the most commonly-used approximations in nuclear reaction networks, it is straightforward to create other approximations using the same framework in \pynucastro.

Another easy approximation that \pynucastro\ supports is modifying a rate.  In
this case, the rate itself is unchanged, but the endpoint nuclei are, and the $Q$ value for the rate is recomputed.  An example use case for this is if you want to capture carbon burning in a network.  \reaclib\ has 3 rates that involve $\isotm{C}{12} + \isotm{C}{12}$, but if you created your network without neutrons,
then you will not have pulled in $\isotm{C}{12}(\isotm{C}{12},\neutron)\isotm{Mg}{23}(\neutron,\gamma)\isotm{Mg}{24}$.  The neutron capture here tends to be very fast, so we could simply approximate this rate as $\isotm{C}{12}(\isotm{C}{12},\gamma)\isotm{Mg}{24}$.  This can be done using the {\tt modify\_products()} method on the rate.  The end result is that we approximate this branch of carbon burning without having to pull two more nuclei into our network ($\neutron$ and $\isotm{Mg}{23}$).  This was used in the network described in \citet{castro_simplified_sdc},
and is illustrated in the network described in Appendix~\ref{app:subch_approx}.

\section{Tabular Weak Rates}
\label{sec:tabular}

Weak rates are of great interest to many applications, especially in stellar environments. These rates often depend on both temperature and density, requiring the rates to be tabulated prior to integration of a network. \pynucastro\ natively includes tabular rates for sd-shell nuclei pairs with $A$ of 17--28 from \cite{suzuki2016}. These rates are accessed through the \tabularrate\ class and can be included in networks with the more standard \reaclib\ rates.

\begin{figure*}[t]
\centering
\plottwo{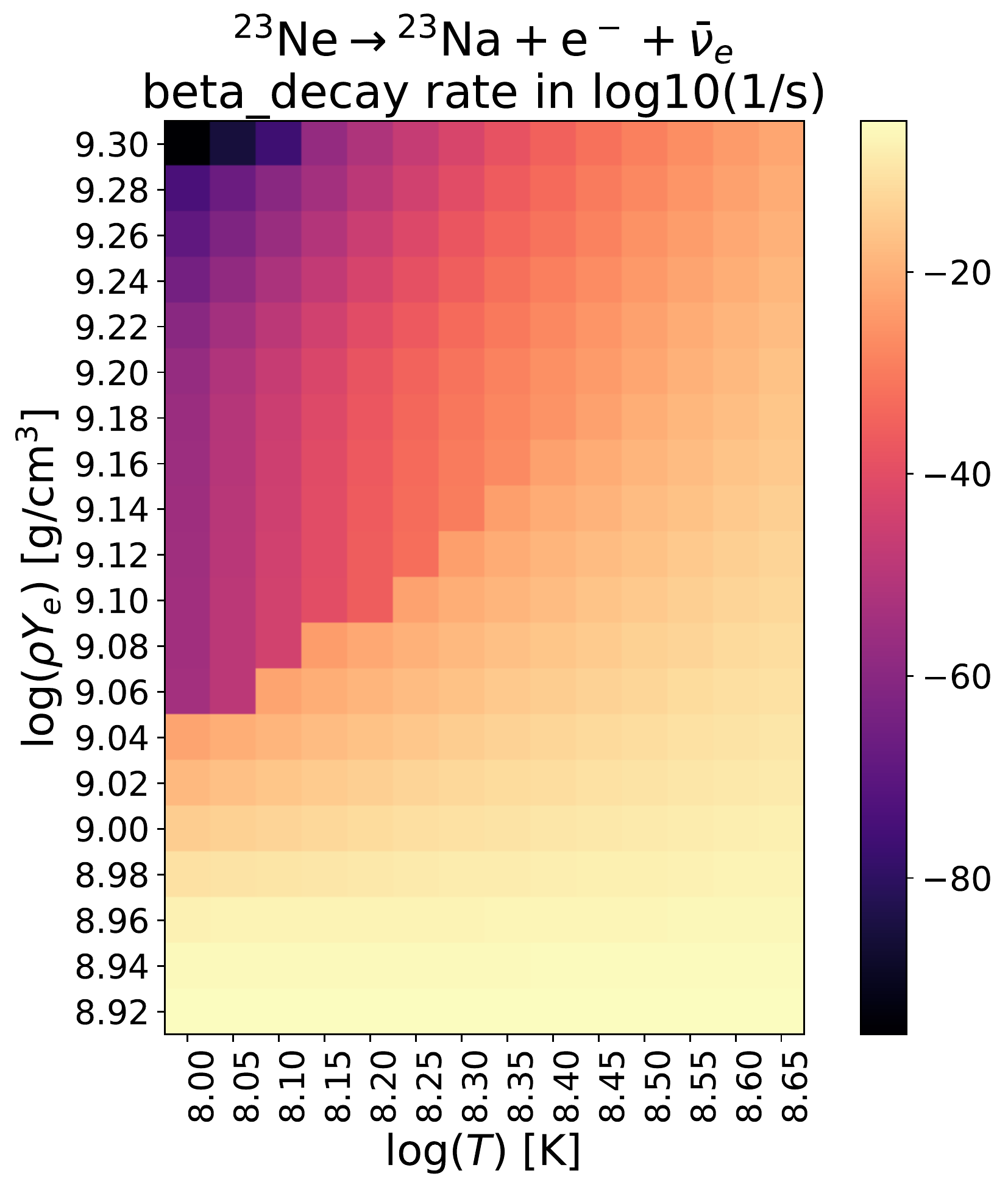}{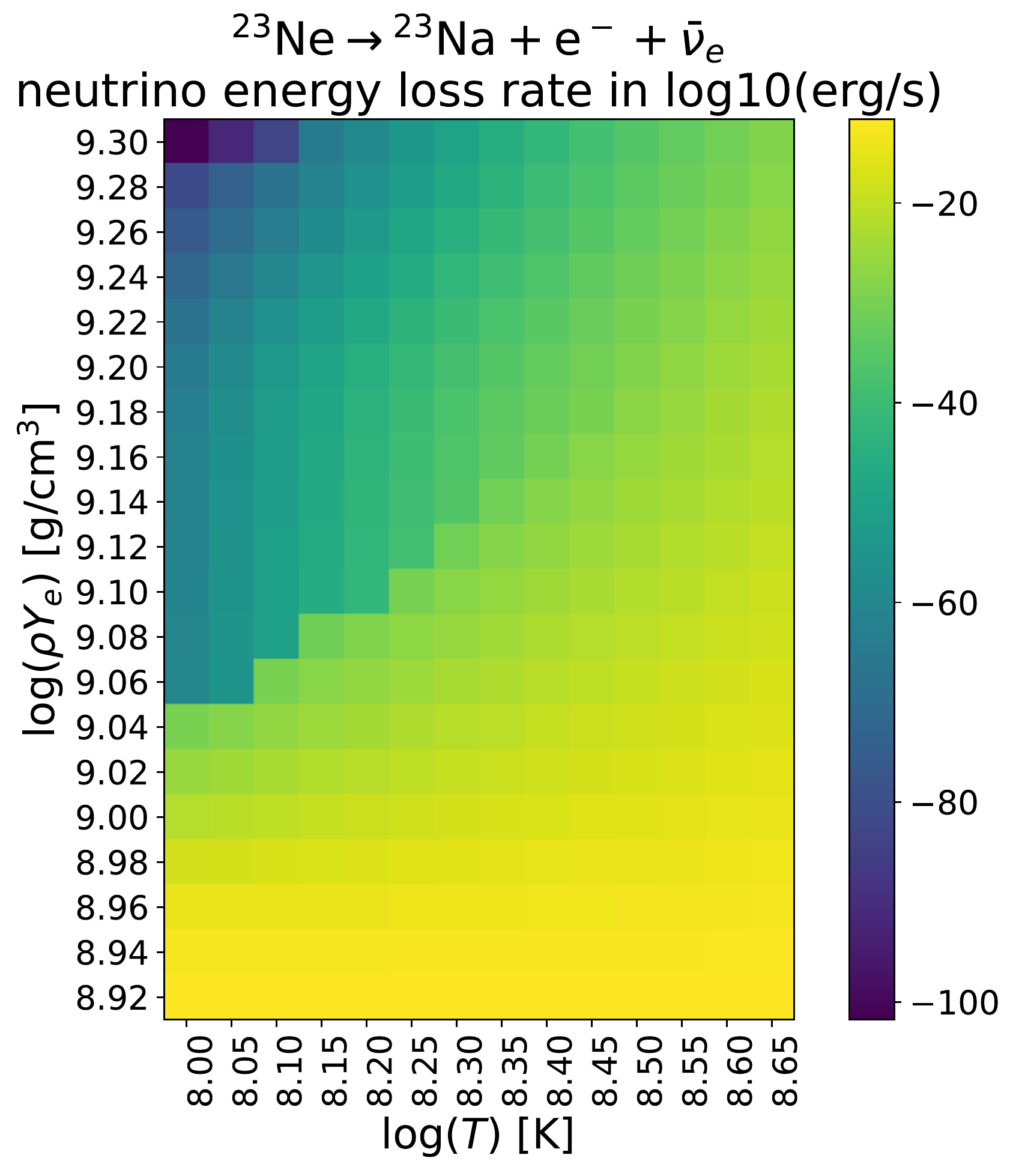}
\caption{\label{fig:weakrate} The left plot shows the reaction rate for the \isot{Ne}{23} beta decay for a range of electron density and temperature. The right plot shows the energy loss rate due to neutrinos for the \isot{Ne}{23} beta decay. \supnote{urca\_network.ipynb}}
\end{figure*}

\tabularrate\ reads a file given in a ($\rho Y_e$, T) format and performs linear interpolation to calculate the appropriate rate and the energy lost to neutrinos. The energy loss presupposes that the neutrinos free stream out of the environment. Figure~\ref{fig:weakrate} demonstrates this capability over a range of electron density and temperature for the \isot{Ne}{23} beta decay reaction. These two plots are both easily made with \tabularrate\ class and allows for exploration of the various weak rates in \pynucastro .

\begin{figure}[t]
\plotone{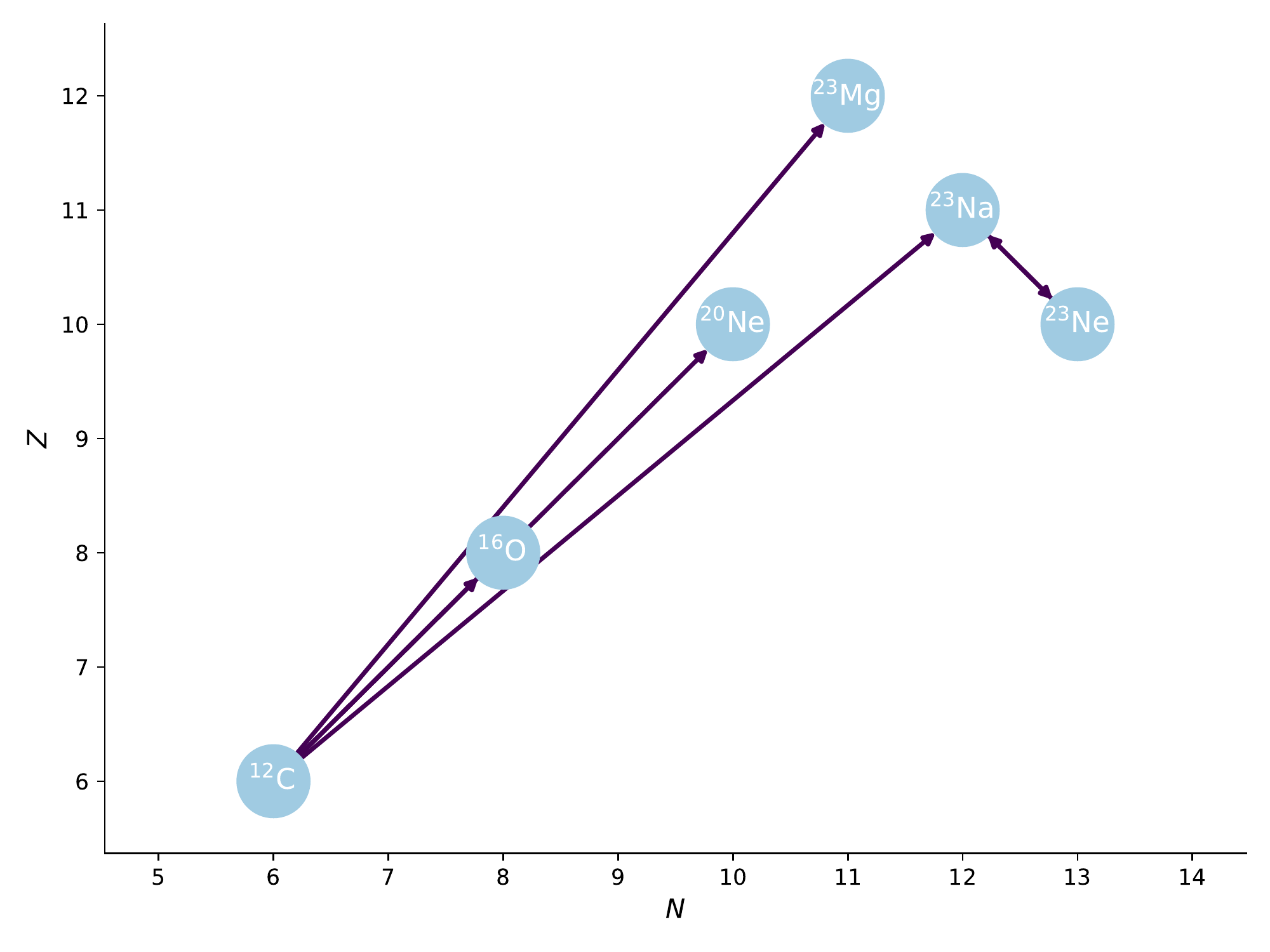}
\caption{\label{fig:urca} A network with simple carbon burning and the $A=23$ Urca reactions. \supnote{urca\_network.ipynb}}
\end{figure}

A nice example of a full network that incorporates \tabularrate\ is the $A=23$ Urca reactions with simple carbon burning. The $A=23$ Urca reactions consist of a beta decay and electron capture, linking the isotopes \isot{Na}{23} and \isot{Ne}{23}.
\begin{equation} \label{eq:urca}
\begin{split}
     \isotm{Ne}{23}           &\rightarrow \: \isotm{Na}{23} + {e}^{-} + {\Bar{\nu}}_{e} \\
     \isotm{Na}{23} + {e}^{-} &\rightarrow \: \isotm{Ne}{23} + {\nu}_{e}
\end{split}
\end{equation}
Both weak reactions were tabulated rates by \cite{suzuki2016}. Rates related to carbon burning are from \reaclib .
Here we see how to create this network with \pynucastro .  
First, we search the \tabularlibrary\ for rates linking \isot{Na}{23} and \isot{Ne}{23}. Then we 
specify rates from the \reacliblibrary\ by using each rate's name:
\begin{lstlisting}
rl = pyna.ReacLibLibrary()
tl = pyna.TabularLibrary()

# get Tabular rates
urca23_nuc = ["na23", "ne23"]
urca23_lib = tl.linking_nuclei(urca23_nuc)
urca23_rates = urca23_lib.get_rates()

# get ReacLib rates
rl_names = ["c12(a,g)o16",
            "c12(c12,a)ne20",
            "c12(c12,p)na23",
            "c12(c12,n)mg23",
            "n(,e)p",]
rl_rates = rl.get_rate_by_name(rl_names)

#combine rates in one network
rates = urca23_rates + rl_rates
urca23_net = pyna.PythonNetwork(rates=rates)
\end{lstlisting}
The resulting network is shown in Figure~\ref{fig:urca}.
This demonstrates how we can easily incorporate both \tabularrate\ and \reaclibrate\ rates into a single network. 
For \cxx\ networks, \amrexastrocxxnetwork\ automatically adds the calls to interpolate from the rate tables when
filling the reaction rates.
This network was used for convective Urca simulations \citep{urca} with the \maestroex\ code \citep{maestroex}.  

\section{Reverse Rates and Nuclear Partition Functions}
\label{sec:actual_detailed_balance}

The relationship between the inverse and the forward reaction may be obtained by detailed balance calculations---under the assumption of equilibrium in the reaction, we can derive the inverse reaction rate, $N_a^{p-1}\langle \sigma v\rangle_{B_1,B_2,\cdots,B_p}$ in terms of the forward rate $N_a^{r-1}\langle \sigma v\rangle_{C_1,C_2,\cdots,C_r}$ (see the derivation in Appendix \ref{sec:detailed_balance}). Each detailed balance inverse rate can be decomposed as a multiplication between a component that satisfies the fit (\ref{eq:reaclib_fit}), as in \reaclib, and an additional component that involves partition functions:
\begin{align} \label{eq:det_balance}
     N_a^{p-1}\langle \sigma v \rangle_{B_1,B_2\cdots,B_p} &=  N_a^{p-1}\langle \sigma v \rangle_{C_1,C_2\cdots,C_p}' \nonumber \\ 
     &\hspace{0.5cm}\times\, \dfrac{G_{C_1} \cdots G_{C_r}(T)}{G_{B_1} \cdots G_{B_p}(T)},
\end{align}
where $N_a^{p-1}\langle \sigma v \rangle_{B_1,B_2\cdots,B_p}$ is defined by (\ref{eq:reverse}), and $N_a^{p-1}\langle \sigma v \rangle_{B_1,B_2\cdots,B_p}'$ is the rate parametrization (\ref{eq:reaclib_fit}) with the constants $a_{0, \mathrm{rev}}$, $a_{1, \mathrm{rev}}$, $\cdots$, $a_{6, \mathrm{rev}}$ given by (\ref{eq:a0_rev}--\ref{eq:a6_rev}), and each $G_{n_i}$ is the nuclear partition function of the $n_i$-nuclei defined in (\ref{eq:pf}).

The nuclear partition function information tables are stored in a suitable format to be read by \pynucastro, switching between merging the low and high temperature tables in \citet{rauscher:1997} and \citet{rauscher:2003}, or to use the low temperature table only, in \citet{rauscher:1997}. Furthermore, we can decide between using the finite-range droplet model (FDRM) or the Thomas-Fermi approach with Strutinski integral (ETFSI-Q) mass models to compute the partition function values. In our implementation of the partition functions in \pynucastro, we have considered merging both tables and the FDRM mass model as our default options.

Similarly, the tabulated spin of the ground-state nuclei from \citet{huang:2021} and \citet{wang:2021} is also stored in a suitable format to be read by \pynucastro. From this point, we can switch between using all the unique-value ground-state spin measurements, regardless of the experimental or theoretical arguments, or to filter by keeping only the strong experimental observed measurements. In our implementation for the spin of the ground-state nuclei in \pynucastro, we have considered the latter choice.

In order to compute the inverse rate by detailed balance (\ref{eq:det_balance}), we have implemented the \derivedrate\ class. The \derivedrate\ class inherits from \rate\ and computes the \reaclib\ component that satisfies the fit (\ref{eq:reaclib_fit}) $N_a^{p-1}\langle \sigma v \rangle_{B_1,B_2\cdots,B_p}'$ from the chosen forward rate. However, the \derivedrate\ class may go further and compute (\ref{eq:det_balance}) by using interpolations of the tabulated values of the partition functions as requested. It is important to note that \derivedrate\ does not support constructing inverse weak rate reactions. This decision has been made because weak rates are not required to be in equilibrium to occur (see, for example, \citealt{Clifford-and-tayler, SEITENZAHL200996}), breaking the necessary assumptions to compute (\ref{eq:reverse}).

\begin{figure}[t]
\plotone{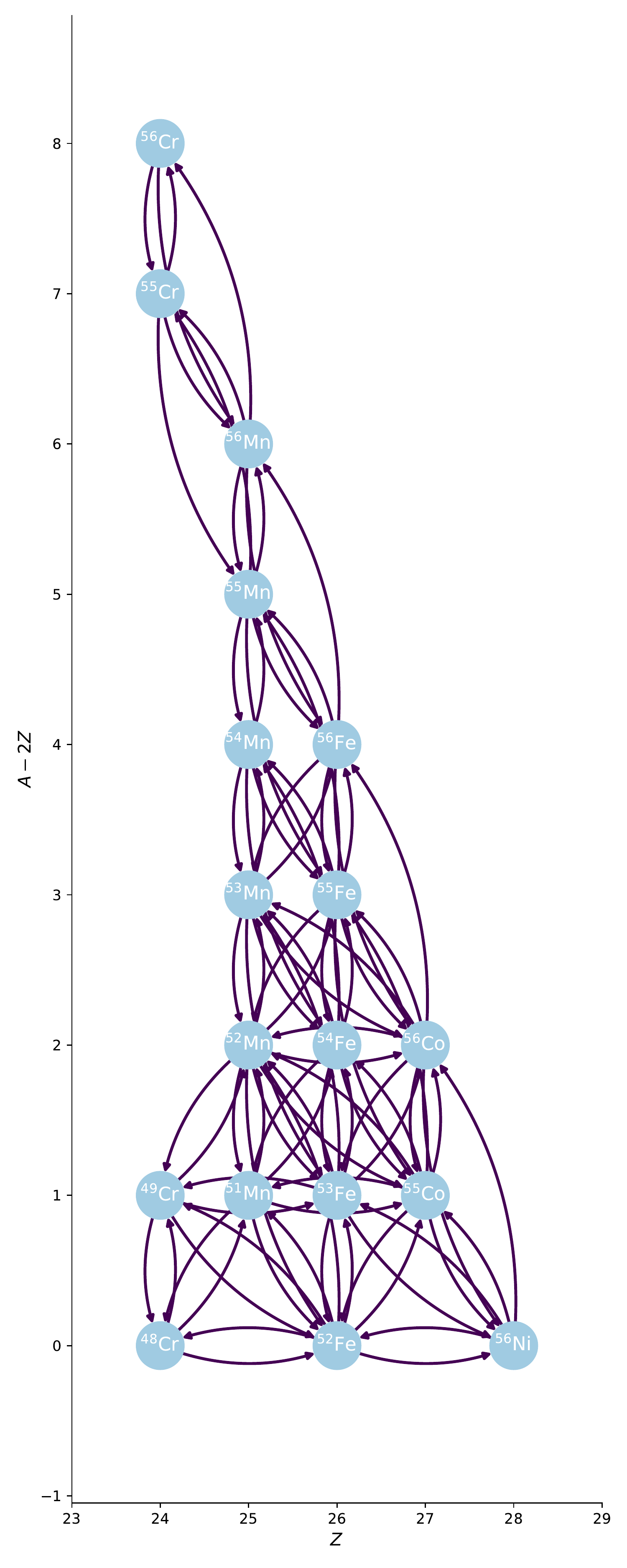}
\caption{\label{fig:derived_network} The integrated network {\tt iron56\_end} inspired by the high-Z portion of the popular {\tt
  aprox21} network by replacing the \reaclib\ detailed balance reverse rates with the \derivedrate\ rates under {\tt use\_pf=True}. \supnote{derived\_network.ipynb}} 
\end{figure}

To show the effects of the partition function correction and the
accuracy of the \derivedrate\ rates, we have defined the {\tt iron56\_end} network with
\isot{Cr}{48,49,55,56}, \isot{Mn}{52\mbox{--}56},
\isot{Fe}{52\mbox{--}56}, \isot{Co}{55, 56}, \isot{Ni}{56} species, along with
$\mathrm{n}$, $\mathrm{p}$, and $\alpha$.  This network is inspired by the high-Z portion of the popular {\tt
  aprox21} network, as illustrated in Figure \ref{fig:derived_network}.

The required code to create {\tt iron56\_end} is:
\begin{lstlisting}
rl = pyna.ReacLibLibrary()
fwd = rl.derived_forward()

nuclei = ["n", "p", "he4",
          "cr48", "cr49", "cr55",
          "cr56", "mn51", "mn52",
          "mn53", "mn54", "mn55",
          "mn56", "fe52", "fe53", 
          "fe54", "fe55", "fe56",
          "co55", "co56", "ni56"]

frates = fwd.linking_nuclei(nuclist=nuclei,
                            with_reverse=False)
derived = []
for r in frates.get_rates():
    d = DerivedRate(rate=r, compute_Q=False,
                    use_pf=True)
    derived.append(d)

der_rates = Library(rates=derived)
full_library = frates + der_rates

pynet = PythonNetwork(libraries=full_library,
                      do_screening=True)
\end{lstlisting}
The screening and inverse screening factors of the {\tt iron56\_end} network are computed from (\ref{eq:f_factor}) and (\ref{eq:f_factor_inv}) respectively. It is important to mention that (\ref{eq:f_factor_inv}) is computed under the assumption of the linear mixing rule, but not all the screening routines may satisfy this requirement (see, for example, \citealt{wallace:1982}). In order to cover these cases, we have implemented symmetric screening to the forward and inverse rate, i.e. the same enhancement factor assigned to the forward rate is assigned also to the inverse rate. 

\begin{figure*}[t]
\centering
\plottwo{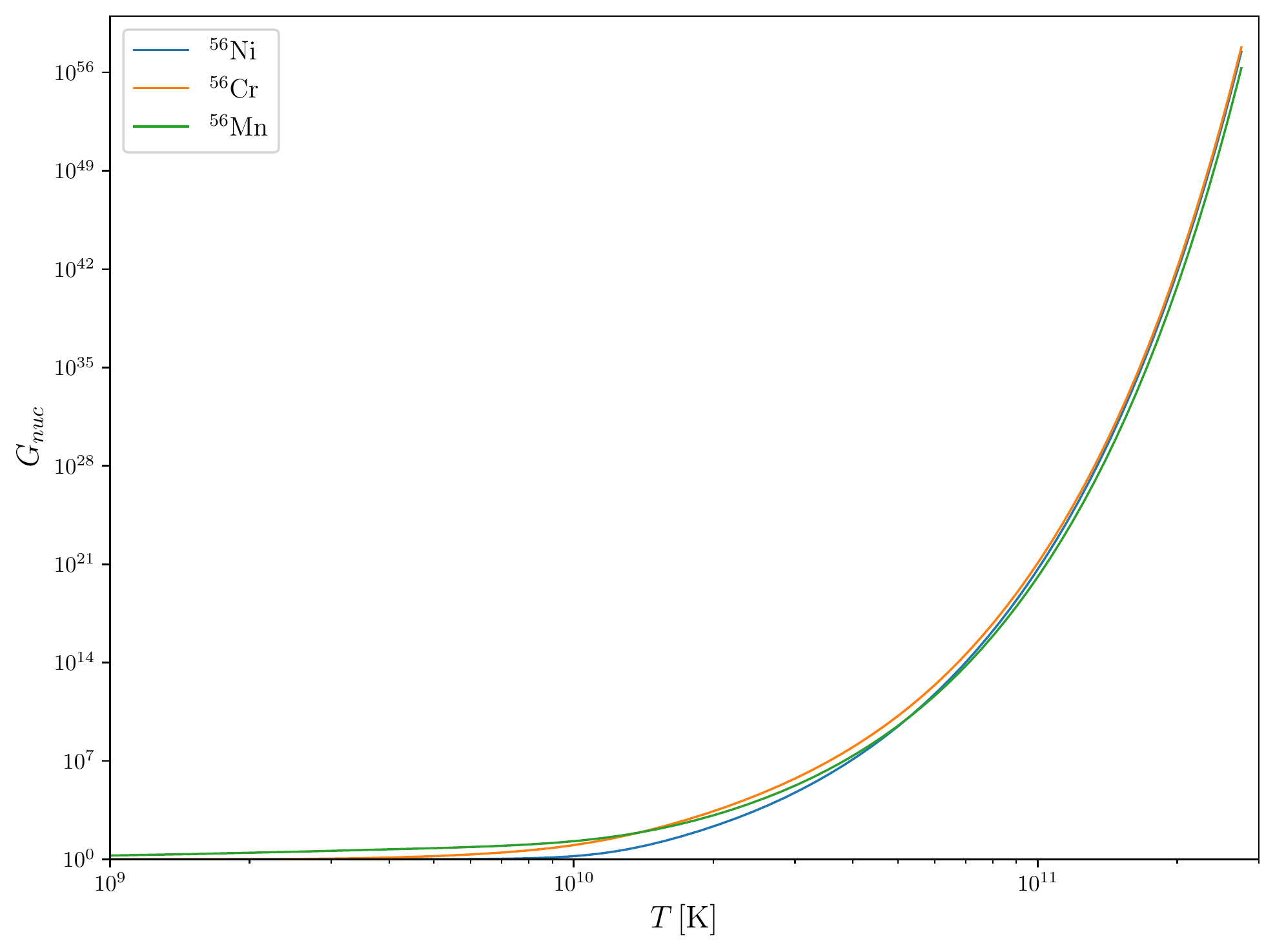}{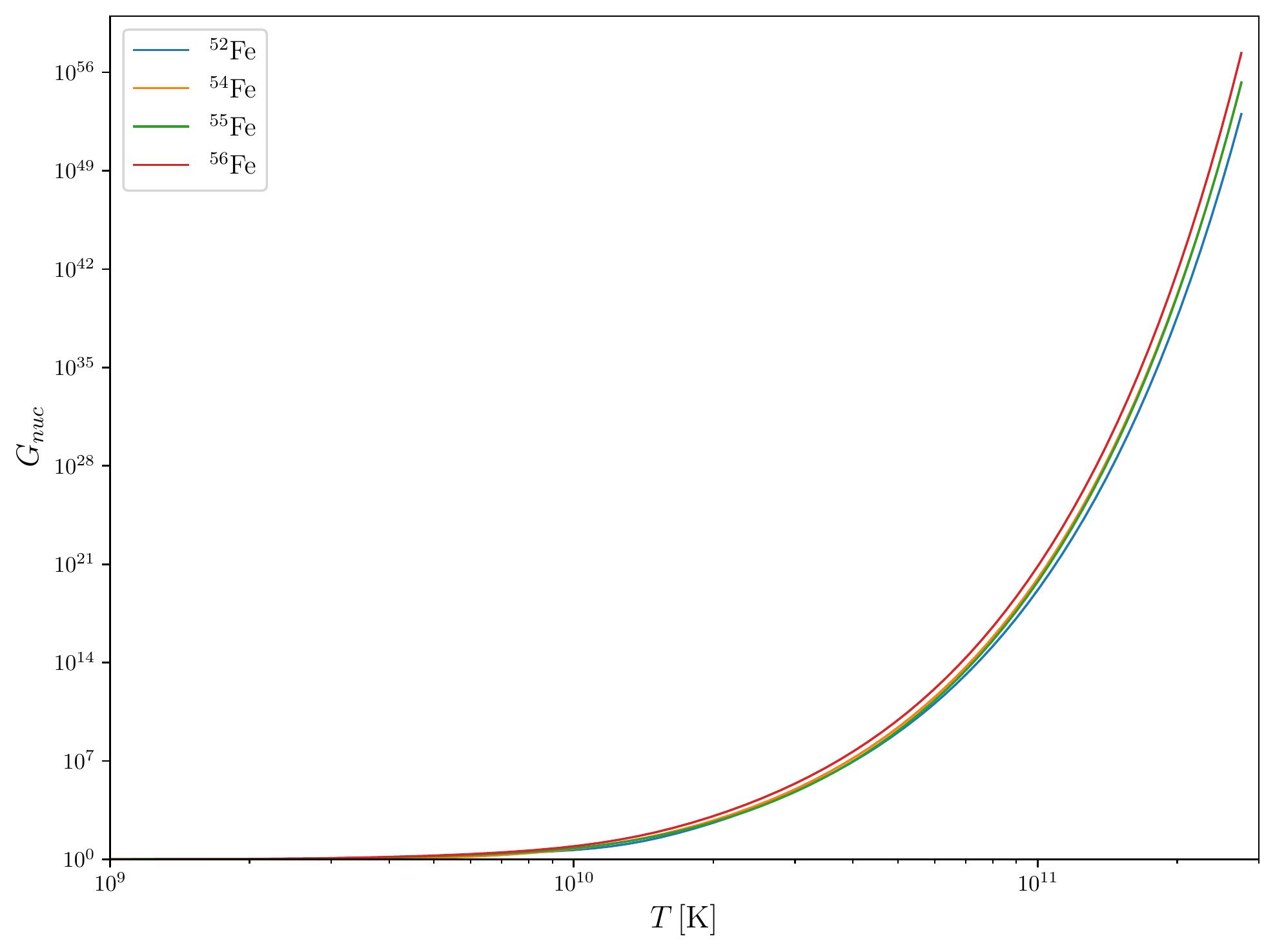}
\caption{\label{fig:pf} An example of the partition function temperature behaviour of \isot{Ni}{56}, \isot{Cr}{56}, \isot{Mn}{56} (left), and \isot{Fe}{52}, \isot{Fe}{54}, \isot{Fe}{55}, \isot{Fe}{56} (right). The partition function value of each nuclei, with identical atomic mass number $A$ but different atomic number $Z$, changes significantly around  $~ 10^{10} \, \mathrm{K}$. This behaviour is different in nuclei with different $A$ and same $Z$, where all the nuclei tends to similar values around  $~ 10^{10} \, \mathrm{K}$. Although we should not generalize this behaviour to all existing nuclei from this particular case, this behaviour dominates in the nuclei of our network. \supnote{plot\_pf.ipynb} }
\end{figure*}

We can explore the partition function temperature behaviour by plotting $G_{n_i} = G_{n_i}(T)$ for different nuclei $n_i$ of the network, as depicted in Figure   \ref{fig:pf}. Notice that the nuclei with the same atomic number have a similar behaviour than the nuclei with the same atomic mass number. 

\begin{figure*}[t]
\centering
\epsscale{1.1}
\plottwo{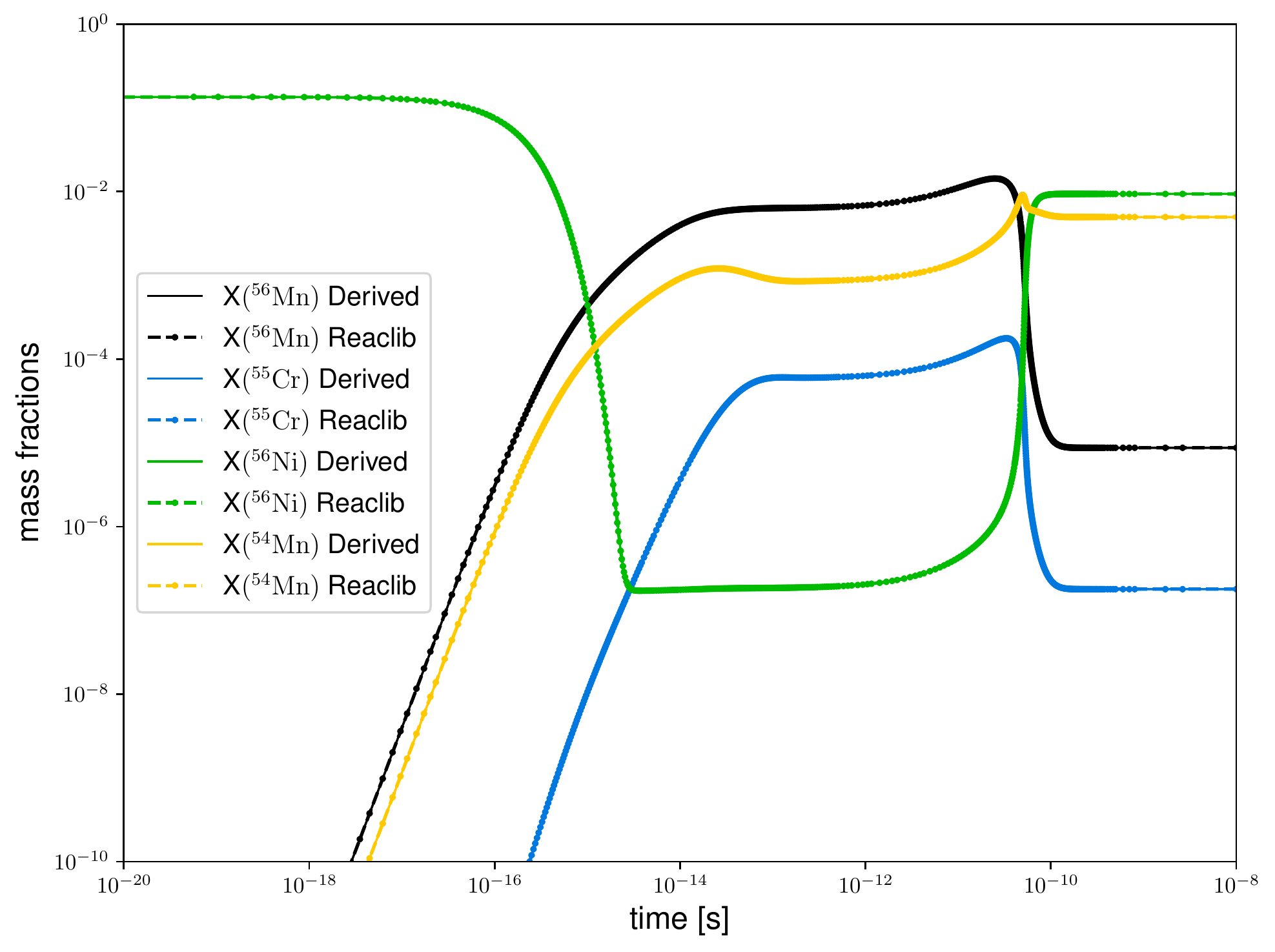}{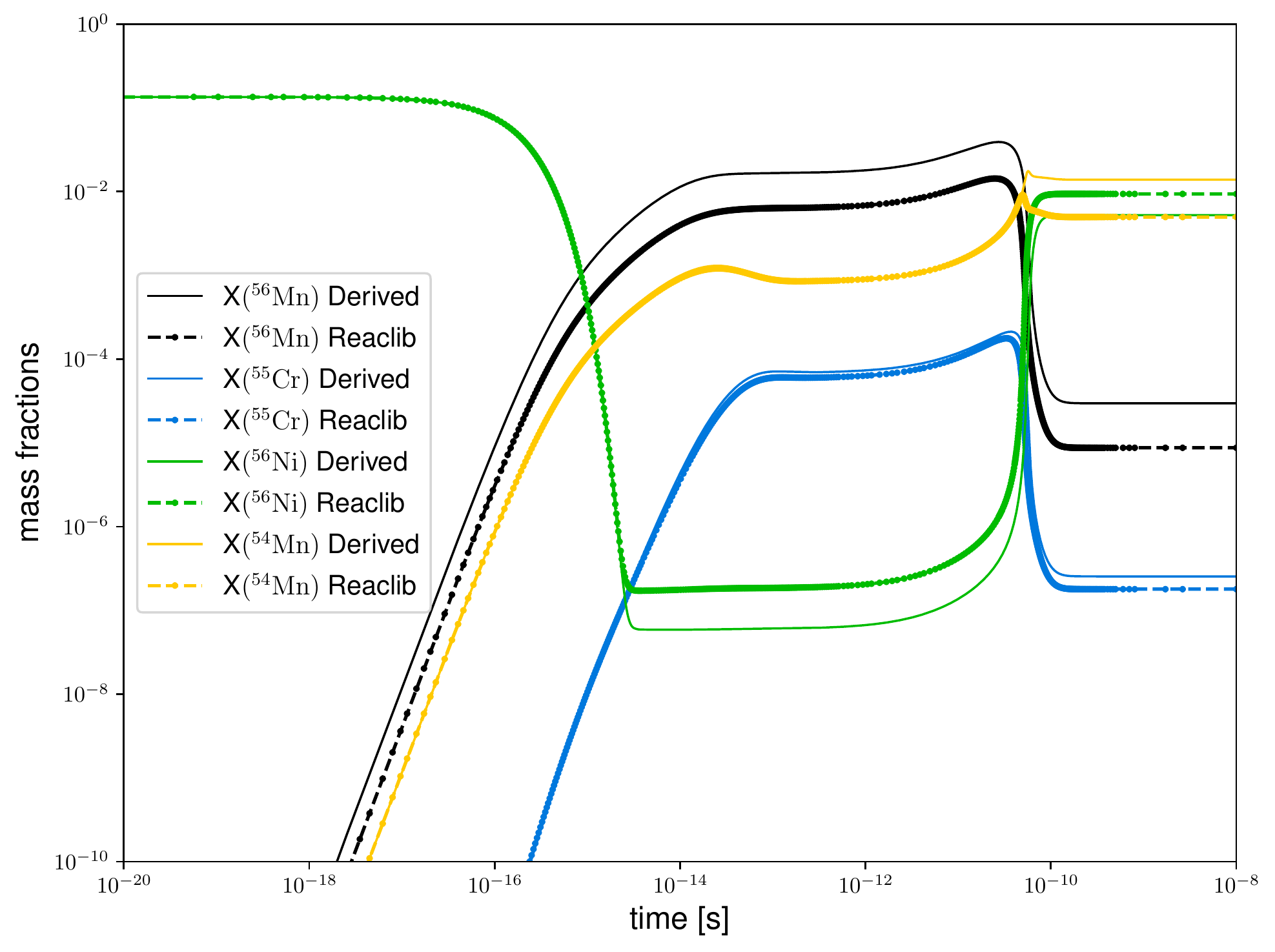}
\caption{\label{fig:network_comparison} Comparing the nuclei $\isotm{Cr}{55}$, $\isotm{Mn}{54,56}$,  and $\isotm{Ni}{56}$ of the {\tt iron56\_end} network with the \reaclib\ detailed balanced inverse rates replaced by \derivedrate\ objects under the effects of the partition functions (right) and the {\tt iron56\_end} network comparison without any modification from the partition functions (left). The maximum molar fractions absolute error in the left case is $\sim10^{-7}$. The density and temperature in both comparisons are  $\rho=10^{8}~\gcc$ and $T=7.0 \times 10^{9}\; \mathrm{K}$ respectively. \supnote{iron56\_end\_networks.ipynb}} 
\end{figure*}

To validate our results, we have compared the results obtained from the {\tt iron56\_end} network by replacing the \reaclib\ detailed balance rates with \derivedrate\ rates, and the results obtained by integrating the {\tt iron56\_end} network with just \reaclib\ detailed balance inverse rates. Setting $\rho=10^{8}~\gcc$, $T=7.0 \times 10^{9}\; \mathrm{K}$, and the composition:

\begin{subequations}
\begin{align}
    X(\proton) = X(\neutron) = X(\alpha) = X(\isotm{Ni}{56}) &=5/37,  \label{eq:network_init_1}\\
    X(\isotm{Cr}{48}) &= 17/37, \label{eq:network_init_4}
\end{align}
\end{subequations}
with all the remaining mass fractions set to zero as our initial conditions, we can show the comparison of these two reaction network calculations and introduce the role of the partition function in the previous reaction network under the same initial conditions (Figure \ref{fig:network_comparison}). The convergence of the network with the inclusion of partition function may differ significantly as we increase the temperature around $\sim10^{10}\;\mathrm{K}$. This temperature regime is common in stellar atmospheres of neutron stars or inside the cores of white dwarfs, where the nuclear statistical equilibrium regime takes place. 

\section{Electron Screening} 
\label{sec:screen}

So far we have been treating the nuclei as if they were decoupled from the surrounding sea of electrons, but inside a star, densities are sufficiently high to invalidate this approximation. Accurate reaction rate calculation requires us to account for Coulomb screening, where the free electrons surrounding each nucleus are preferentially attracted to the positively charged nuclei. This reduces the
effective charge of nuclei, thus lowering the Coulomb barrier to fusion,
and greatly enhancing reaction rates.
From \citet{dewitt:1973} and imposing the linear mixing rule, any reaction rate with a set of reactants $\{C_i\}$ and a set of products $\{B_j\}$:
\begin{equation}
    C_1 + C_2 + \cdots + C_r  \rightarrow B_1 + B_2 + \cdots + B_p
\end{equation}
is enhanced due to the Coulomb screening of the reactants, by a factor $f$ given by:
\begin{align}
    f_{C_1 \cdots C_p} &= \exp\left[\dfrac{\mu^C(Z_{C_1}) + \cdots + \mu^C(Z_{C_r})}{\kb T} \right. \nonumber\\
    &\hspace{1.5cm}\left. - \dfrac{\mu^C(Z_{C_1} + \cdots Z_{C_r})}{\kb T} \right] \label{eq:f_factor}
\end{align}
Next, we need to understand the functional form of $\mu^C_i$, which appears as a part of the free energy $F^{sc}=\mu^C/\kb T$. In \pynucastro\ we have implemented three different fits that construct $F$ and perform the enhancement factor calculations in a reaction network: {\tt chugunov\_2007} from \citet{chugunov:2007}, {\tt chugunov\_2009} from \citet{chugunov:2009}, and {\tt potekhin\_1998} from \citet{potekhin}.  Note
that only \citet{potekhin} is available to be used in the NSE state that we discuss in the next section. Since this fit is implemented in both reactions networks and NSE, we will center our discussion on it.  

We start by defining the $i$-th nuclei Coulomb-coupling constant $\Gamma_i$ of a multi-component plasma (MCP) by:
\begin{equation}\label{eq:sc_def}
    \Gamma_i = \Gamma_e Z_i^{5/3}, \hspace{0.2cm} \Gamma_e = \dfrac{e^2}{a_e\kb T}, \hspace{0.2cm} a_e = \left(\dfrac{4\pi n_e}{3}\right)^{-1/3}
\end{equation}
The approximation $n_e \sim \rho Y_e/m_u$ is used in some references, for example, in \citet{SEITENZAHL200996}, however, in \pynucastro\ we always update $Y_e$ to be consistent with the current estimate of the composition.  Two important containers classes store the necessary information to compute the screening factor: {\tt PlasmaState} and {\tt ScreenFactors}. A {\tt PlasmaState} object holds the temperature, density, the molar fraction and the atomic number values of each nuclei inside the reaction network to be screened. Similarly, a {\tt ScreenFactors} object holds the reactant nuclei information of the rate to be screened. Given the two nuclei contained in a {\tt ScreeningPair} object, a {\tt ScreenFactors} object is initialized to hold the two-reactant nuclei information. 

According to \citet{potekhin} and the definitions provided in (\ref{eq:sc_def}):
\begin{align}
F^{sc}_i &= \dfrac{\mu^C_i}{\kb T} = A_1 \left[\sqrt{\Gamma_i(A_2 + \Gamma_i)} \right. \nonumber\\ 
&\hspace{0.5cm}  -  A_2 \left.\ln{\left(\sqrt{\frac{\Gamma_i}{A_2}} \sqrt{1+\frac{\Gamma_i}{A_2}} \right)} \right] \nonumber\\ 
&\hspace{0.5cm} + 2A_3\left[\sqrt{\Gamma_i} - \arctan(\sqrt{\Gamma_i})\right]   \label{eq:coulomb_corr}
\end{align}
where  $A_1 = -0.9052$, $A_2 = 0.6322$, and $A_3 = -\sqrt{3}/2 - A_1/\sqrt{A_2}$ is valid within a range $0.01 < \Gamma_i \lesssim 170$. The {\tt iron56\_end} network, under the initial conditions (\ref{eq:network_init_1} - \ref{eq:network_init_4}), at $\rho = 10^8\;\mathrm{g}\, \mathrm{cm}^{-3}$ and $T=7.0 \times 10^{9}\; \mathrm{K}$, defines a range $0.01 < \Gamma_i <3.09$, which agrees within the boundaries of the fit.  This fit is provided in \pynucastro\ by the function {\tt potekhin\_1998}.

Once the enhancement calculations are completed and returned to the {\tt RateCollection} network, we rely on the {\tt ScreeningPair} class and the {\tt evaluate\_screening} method to hold the necessary information to compute the screening factor. The {\tt ScreeningPair} class is a container of all the same two-nuclei reactant rates. The core idea of this class is to contain the required information, relative to the reactants. For \reaclib\ we use only one {\tt ScreeningPair} object for each reaction, with the exception of the triple-$\alpha$ reaction. The method {\tt evaluate\_screening} constructs a {\tt PlasmaState} object from the density, temperature, and molar composition of the network, to compute the $n_e$ factor, $\Gamma_e$, and $\Gamma_i$ respectively. It then creates a list with all the {\tt ScreeningPair} objects required to screen the network, holding their information inside {\tt ScreenFactors}. Finally, it uses these objects to compute the screening factor for each rate using the specified fit method. It is important to recall that we have described only the use of the {\tt potekhin\_1998} method, but {\tt chugunov\_2007} and {\tt chugunov\_2009} are also available in our implementation. 

Finally, we note that it is easy to use one of the screening
implementations when integrating a python network---simply
add the name of the screening function to the {\tt args} tuple
discussed in section \ref{sec:integrating} and it will be called
when evaluating the rates.  For \cxx\ networks, we have \cxx\ ports
of the screening routines that are automatically hooked in when
using \amrexastrocxxnetwork.

\section{Nuclear Statistical Equilibrium}
\label{sec:nse}

One additional feature we have implemented in \pynucastro, is the
ability to compute the nuclear statistical equilibrium (NSE) mass
fractions using several iterations of the hybrid Powell's method \citep{powell:1964,powell:1970}.  In
python, this is implemented by the SciPy {\tt fsolve()} function.  In
\cxx, we converted the Fortran 77 implementation of Powell's method
from MINPACK \citep{minpack} to \cxx\ as a templated header library.
This solver is more robust than the traditional Newton-Raphson method
in two ways. From our tests, Powell's method can handle a larger range
of thermodynamic conditions than simple Newton-Raphson iterations. It
is also capable of handling the problem of the Jacobian becoming singular
during Newton-Raphson iterations.  As pointed out in \citet{skynet},
we will follow the same strategy to accelerate the solution
convergence, by updating $Y_e$ from the computed composition of the previous iteration.

We begin setting the NSE variables $(\rho, T, Y_e)$,  and an initial guess for $(\mu^\mathrm{id}_p, \mu^\mathrm{id}_n)$ -- we have chosen $(\mu^\mathrm{id}_p, \mu^\mathrm{id}_n)$ to be  $(-3.5, -15.0)$, however a different pair may be chosen if the solution does not converge uniformly. We start the first iteration with  $\mu^C_i = 0$.  The solution is then obtained by iterating over the following:
\begin{enumerate}
    \item First, we compute the NSE mass fractions (see the derivation in Appendix \ref{sec:nse_derivation}):
\begin{align}
  X_i &= \dfrac{m_i}{\rho}(2J_i + 1)G(T) \left(\dfrac{m\kb T}{2\pi \hbar^2} \right)^{3/2} \nonumber \\
  &\hspace{0.5cm}\times\exp \left[\dfrac{Z_i \mu_p^{\mathrm{id}} + N_i\mu_n^{\mathrm{id}} + Q_i}{\kb T} \right. \nonumber\\
  &\hspace{2.0cm} \left. - \dfrac{\mu^C_i + Z_i\mu_p^C }{\kb T} \right]  \label{eq:ap_nucleon_fraction}
\end{align}
     
    \item Next we use the predicted mass fractions in the constraint equations:
\begin{subequations}
\begin{align}
    \sum_{i} X_i -1 = 0 \label{eq:nse_constraint_eq1}\\
    Y_e - \sum_{i} \dfrac{Z_i X_i}{A_i} = 0 \label{eq:nse_constraint_eq2}
\end{align}
\end{subequations}
 and solve for $\mu^\mathrm{id}_p$ and $\mu^\mathrm{id}_n$. The first constraint equation ensures that the predicted mass fractions sum to one and the second enforces that the electron fraction of the state agree with the electron fraction input.
    
    \item From the computed $(\mu^\mathrm{id}_p, \mu^\mathrm{id}_n)$ pair we evaluate the network composition of the nuclei, using (\ref{eq:ap_nucleon_fraction}) and the electron number density $n_e$ from:
    \begin{equation}\label{eq:n_e}
        n_e = \dfrac{\rho}{m_u} \sum_{i} \dfrac{Z_i X_i}{A_i}.
    \end{equation}
    where $m_u$ is the atomic unit mass. 
    
    \item Using the equations (\ref{eq:sc_def}) and (\ref{eq:coulomb_corr}), we compute $\mu^C_i$ by replacing the computed electron number density of the previous step (\ref{eq:n_e}). Is important to recall that we may take $\mu^C_i = 0$ for any iteration if no Coulomb screening is considered. 
    
    \item Finally, we compute $\mu^\mathrm{id}_p$ and $\mu^\mathrm{id}_n$, setting each nuclei $\mu^C_i$ to the value computed in the previous step and by the use of the hybrid Powell's method again. 
\end{enumerate}
    
We iterate this process until the values of $\mu^\mathrm{id}_p$ and $\mu^\mathrm{id}_n$ converge to an absolute and relative tolerance of $\sim 10^{-10}$. These values may be adjusted by hand, as the initial guess, to ensure the convergence of the solution.

The method {\tt get\_comp\_nse} implemented in the \ratecollection\ network follows the previous four steps to compute the solution, ensuring that $\mu^C_i$ is constant for each iteration, while the hybrid Powell's method compute the solution. A sample code that solves the NSE state given a set of nuclei with Coulomb correction is shown below: 

\begin{lstlisting}
rl = pyna.ReacLibLibrary()

nuclei = ["n", "p", "he4",
          "c12", "o16", "ne20",
          "mg24", "si28"]

lib = rl.linking_nuclei(nuclei)
rc = pyna.RateCollection(libraries=lib)

rho = 1.0e7   # gram/(cm^3)
T = 6.0e9     # Kelvin
ye = 0.5      # unitless electron fraction

nse_comp = rc.get_comp_nse(rho, T, ye,
                use_coulomb_corr=True)
\end{lstlisting}

\begin{figure}[t]
\centering
\plotone{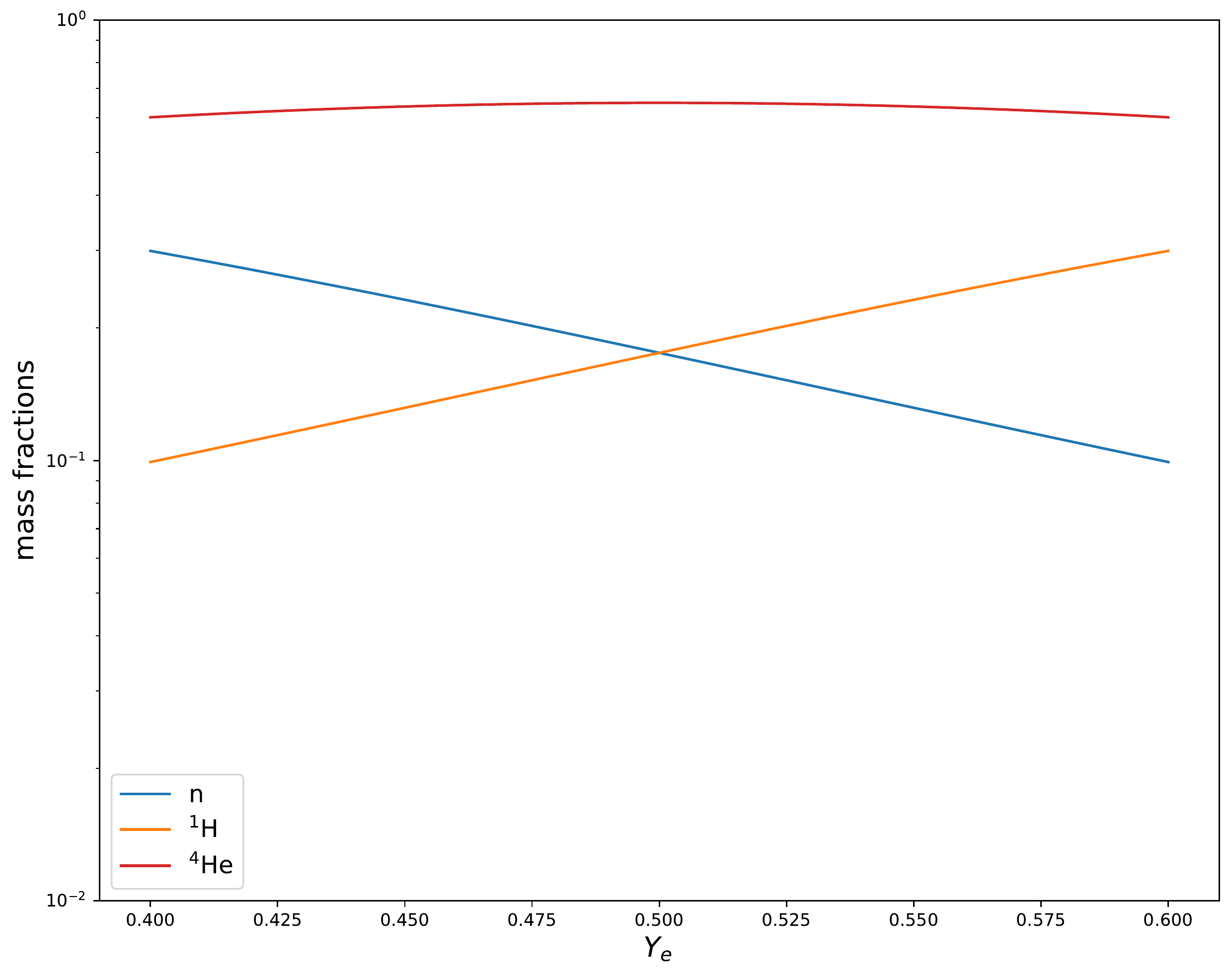}
\caption{\label{fig:nse_fig1} This figure shows the NSE mass fraction at a range of electron fractions from $0.4$ to $0.6$ at $\rho = 10^7 \textrm{g cm}^{-3}$ and at temperature, $T = 9.0 \times 10^9 \; \mathrm{K}$. The nuclei \isot{p}{}, \isot{n}{}, and \isot{He}{4} dominates at high temperature and there is a symmetry correspondence between the mass fraction of \isot{p}{} and \isot{n}{} once $Y_e = 0.5$. A similar figure is observed in Figure 1 in \citet{Seitenzahl_2008}. \supnote{nse.ipynb}}
\end{figure}

\begin{figure}[t]
\centering
\plotone{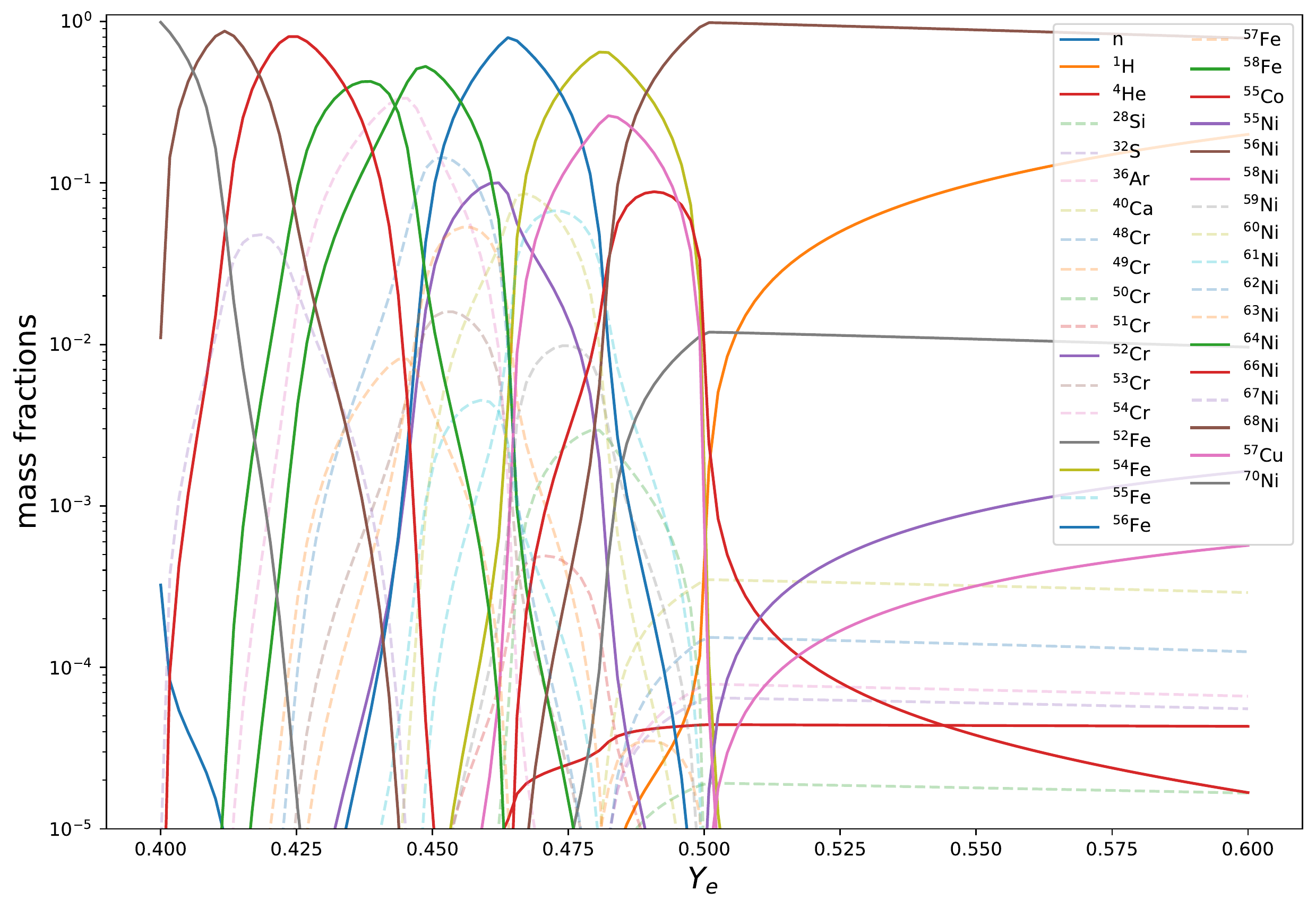}
\caption{\label{fig:nse_fig3} This figure shows the NSE mass fraction at a range of electron fractions from $0.4$ to $0.6$ at $\rho = 10^7 \textrm{g cm}^{-3}$ and at temperature, $T = 3.5 \times 10^9\,\mathrm{K}$. Dashed-lines with transparent colors are nuclei that are not present in Figure 3 in \citet{Seitenzahl_2008} for easier comparison. This figure shows a distinct behavior once $Y_e = 0.5$, which is observed in \citet{Seitenzahl_2008}. \supnote{nse.ipynb}}
\end{figure}

Figure \ref{fig:nse_fig1} and \ref{fig:nse_fig3} are shown to compare the computed NSE distribution at different thermodynamic conditions to the ones obtained from literature to show the validity of the solver. The nuclei that are not present in the corresponding figures for comparison are in dashed-lines and transparent colors for better readability. Both Figures \ref{fig:nse_fig1} and \ref{fig:nse_fig3} show how NSE mass fractions change for a range of electron fractions from 0.4 to 0.6 at a fixed density, $\rho = 10^7~\gcc$, and temperature. Figure \ref{fig:nse_fig1} is at a relatively high temperature, $\textrm{T} = 9.0 \times 10^9$ K. There is a domination of mass fractions by \isot{p}{}, \isot{n}{}, and \isot{\alpha}{} and a symmetry in mass fractions at $Y_e = 0.5$. Figure \ref{fig:nse_fig3} is at a relatively low temperature, $\textrm{T} = 3.5 \times 10^9$ K. Mass fractions change drastically before $Y_e = 0.5$, but become steady and stable once $Y_e > 0.5$. Figure 1 and 3 in \citet{Seitenzahl_2008} used the same thermodynamic conditions, and have a similar result for comparison.

\begin{figure*}[t]
\centering
\plottwo{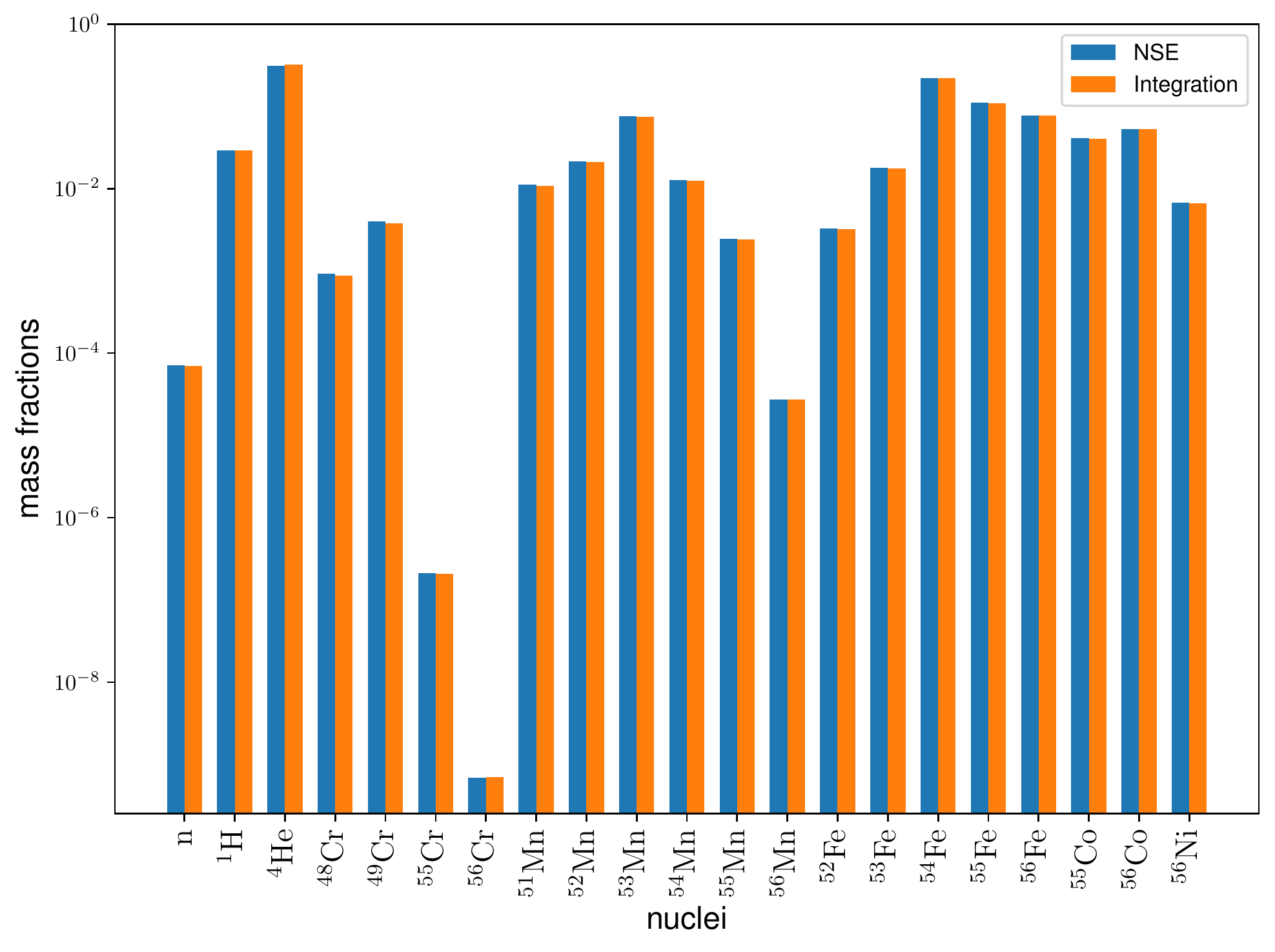}{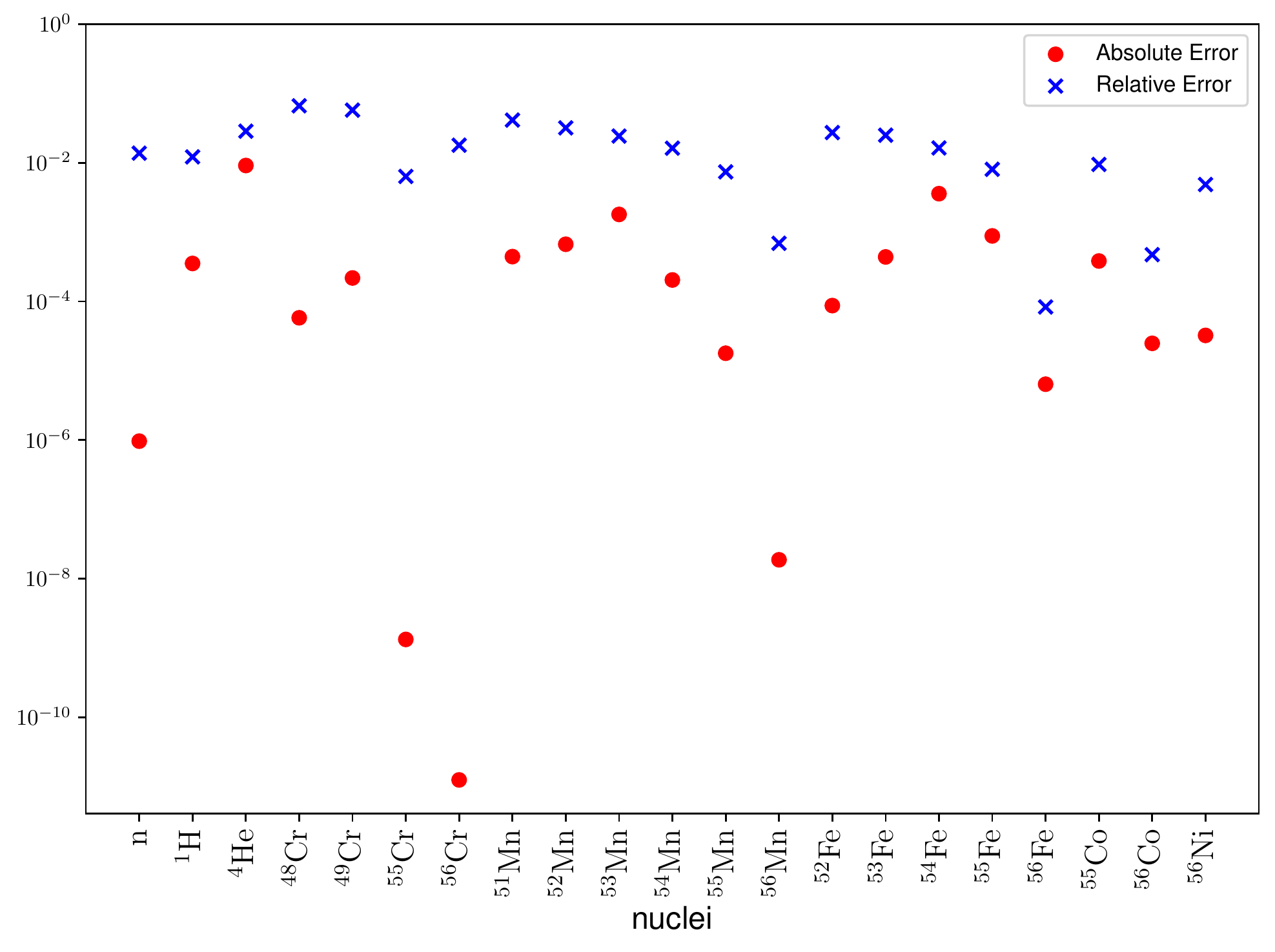}
\caption{\label{fig:nse_vs_network} Comparing the mass fractions $X_i$ of each nuclei $n_i$ contained in the reaction network, by integrating the {\tt iron56\_end} reaction network and computing its NSE mass fractions (left). Also we compute the absolute and relative mass fraction differences with respect to each nuclei (right). The Coulomb correction is included in the NSE state and reaction network calculations. The density and temperature are $\rho=10^{8}~\gcc$ and $T=7.0 \times 10^{9}\; \mathrm{K}$ respectively. \supnote{iron56\_end\_integrate.ipynb}}
\end{figure*}

Finally, in order to compare how the reaction network agrees with the NSE calculations we have considered the {\tt iron56\_end} network, described in Figure \ref{fig:derived_network}, under the initial conditions (\ref{eq:network_init_1}--\ref{eq:network_init_4}). By keeping the default options intact, we obtain the comparison between integrating the reaction network and computing the NSE state mass fractions, shown in Figure \ref{fig:nse_vs_network}. From this comparison we can assert an excellent agreement between the NSE solve and integrating the network to steady state, however it is important to mention that our initial conditions choice significantly impacts the convergence of the steady state of the network into the NSE state composition.

\section{\cxx\ Code Generation}
\label{sec:comparison}

\pynucastro\ works closely with the \amrex-Astro simulation codes,  \castro~\citep{castro_joss} and \maestroex~\citep{maestroex}.  The codes are written in \cxx, for performance portability, and therefore, we need to output reaction networks in \cxx.
A key concern for the \cxx\ code generation is that it be able to work on GPUs as well as CPUs.  We do this in the \amrex-Astro framework, utilizing the \amrex~\citep{amrex_joss} data-structures and lambda-capturing of compute kernels to offload the entire network integration to GPUs. \pynucastro\ generates the righthand side and Jacobian of the network as well as the metadata needed to interpret it, and this can be directly used by our codes.  The \cxx\ networks differ from the simple python networks described above due to the need to couple with hydrodynamics.  \castro\ supports three different coupling strategies: an operator split approach where reactions and advection are independent of one another, a traditional spectral deferred corrections (SDC) approach that supports second- and fourth-order space-and-time integration~\citep{castro_sdc}, and a simplified SDC approach~\citep{castro_simplified_sdc}.  The \cxx\ network generated by \pynucastro\ needs to be compatible with each of these approaches.    As part of this approach, we integrate mass fractions, $X_k$, or partial densities, $(\rho X_k)$, in the reaction network, instead of the molar fractions used internally in \pynucastro.

Integrating astrophysical reaction networks requires stiff
integrators, and a lot of work has been done on assessing
the performance of different algorithms~\citep{timmes_networks,longland:2014}.  The \amrex-Microphysics suite uses the VODE integrator \citep{vode} to integrate the reaction network.  VODE is a 5th order implicit backward-difference integrator that handles stiff reaction networks well.  VODE was originally written in Fortran 77, but we completely ported it to \cxx\ to work in our GPU framework, and added a number of integration step checks to help preserve $\sum_k X_k = 1$ without having to resort to renomalization of the species.  The details of our integrator are given in \cite{castro_simplified_sdc}.

The \amrex-Astro networks are implemented in header files to enable
function inlining.  \pynucastro\ contains templates with these headers
with keyword placeholders for the network-specific code to be injected.  The 
\amrexastrocxxnetwork\ {\tt write\_network()} function reads the templates,
and at each keyword encountered calls an appropriate function to write out
the appropriate piece of the network, for instance functions to evaluate reaction rates, ydot terms, and Jacobian.  We represent
the ydot terms as SymPy objects and use the SymPy \cxx\ code generator to write the expressions to compute them.  Since we evolve energy in the \cxx\ networks, we include temperature derivatives in the Jacobian, and the rate functions use templating to compile this out when not needed (for instance when using a numerical Jacobian).  For our application codes, we generate the networks once, and keep them under version control in our \microphysics\ project repository \citep{microphysics}.  \microphysics\ contains unit tests to exercise any network, which are run nightly.  The \cxx\ networks have the same screening functions as in \pynucastro.  Additionally, the \cxx\ networks have a plasma neutrino loss term, which has not yet been ported to python.  This augments the energy from the network.

\subsection{Comparison of python to \cxx\ nets}

To check for consistency between the python and \cxx\ networks, we do a simple one-zone burn.  Although the \cxx\ networks are often done with hydro evolution, or at least a temperature / energy equation (a self-heating network), we compare without any temperature evolution here. 
We compare using the {\tt subch\_approx} network---this is an approximate version of the network described in detail in \citet{castro_simplified_sdc}.  The code needed to generate it is 
given in Appendix \ref{app:subch_approx}.

In our \cxx\ \microphysics\ \burncell\ unit test, we set the initial thermodynamic state and then instruct VODE to integrate by a $\Delta t$ and return the new solution.  In order to see the history of the evolution, we  evolve to $t = 10^{-3}~\mathrm{s}$ in 100 increments, logarithmically-spaced between $10^{-10}~\mathrm{s}$ and $10^{-3}~\mathrm{s}$.  It is not enough to just disable the energy evolution (by setting $de/dt = 0$, but we also need
to skip obtaining the temperature from the equation of state each step, since even with fixed $e$, the composition change will cause $T$ to change.  With this set, we can compare the \cxx-VODE integration to the python-Scipy {\tt solve\_ivp} integration (we use the {\tt "bdf"} option, which is an implementation of \citealt{matlab-bdf}). 

\begin{figure}[t]
\plotone{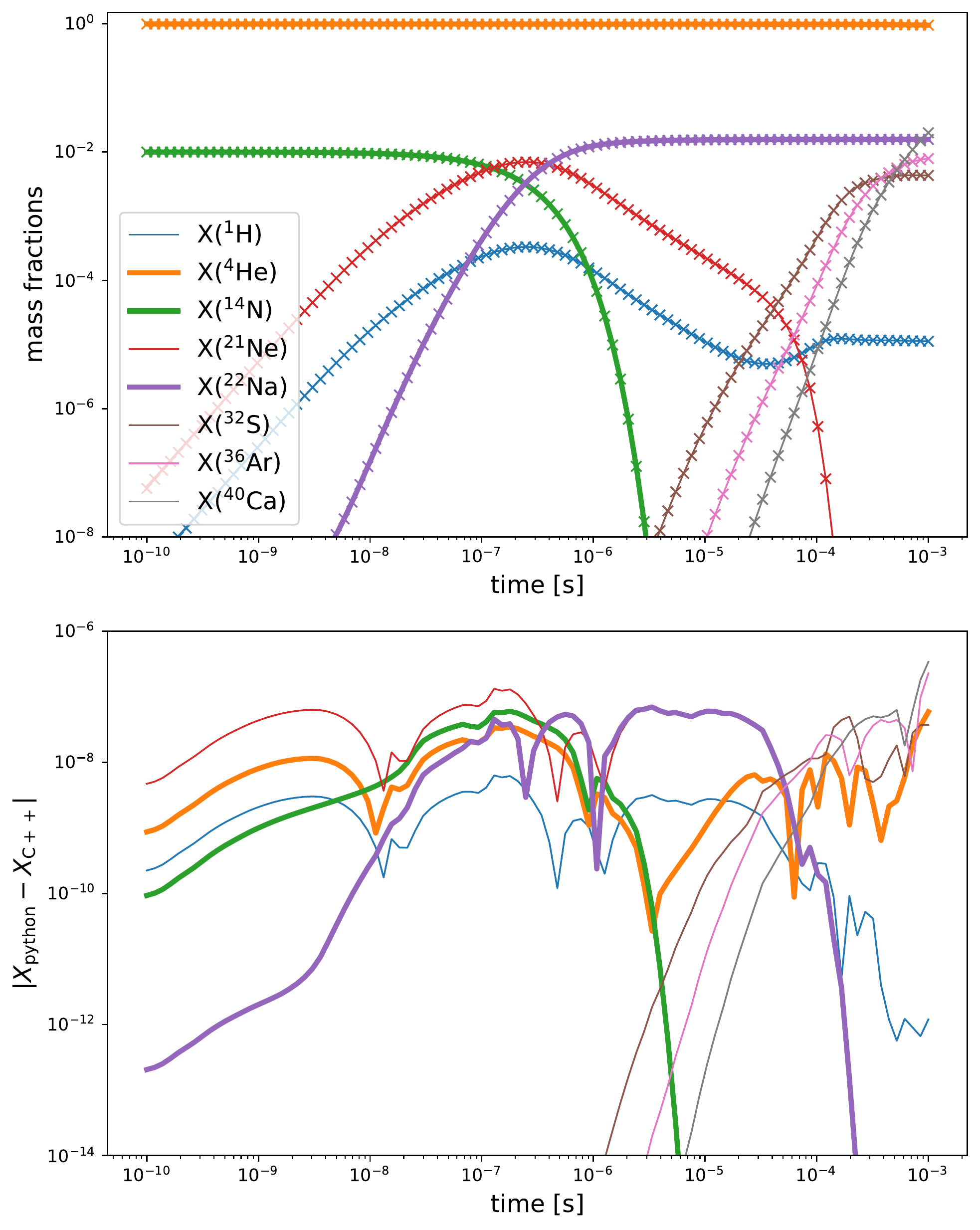}
\caption{\label{fig:cxx_compare} Comparison between the python and \cxx\ integration
of the \subchapprox\ network at fixed $T$ and $\rho$.  The top panel shows the mass fractions, with the python result as lines and the \cxx\ result as points.  The bottom
panel shows the difference between the two
codes. \supnote{subch\_approx\_comparison.ipynb} }
\end{figure}

Figure~\ref{fig:cxx_compare} shows the comparison.  We select $\rho = 10^6~\gcc$, $T = 2\times 10^9~\mathrm{K}$, and the initial composition 99\% \isot{He}{4} and 1\% \isot{N}{14} by mass. On the top plot, we show the history of the mass fractions from python (solid lines) and \cxx\ (points), and see that they lie right on top of one another.  For clarity, we only plots mass fractions that go above $X = 10^{-5}$ at some point in their history.  In the lower panel, for those same species, we plot the difference between the python and \cxx\ abundances as a function of time.  We see that this absolute error is always below $10^{-6}$.  This is despite the integrators being different implementations of a backwards difference integrator. 
  This shows that our \cxx\ and python code generation is remarkably consistent.


\section{Summary and Future Work}

There are many areas of development to focus on in the future.
First, there are a lot of other examples of approximations to
networks that could be added.  For instance, CNO and hot-CNO burning can be reduced to just a few rates, carbon and oxygen burning can likewise be simplified to a few endpoints, and approximations to iron-group equilibrium like those made in the {\tt aprox19} network and others can make lower $Y_e$ accessible with a reduced network.  Adding a separate
``photo-ionization'' proton, like done in {\tt aprox19} is also desirable,
since it reduces the complexity of the Jacobian of the system.  We can build a derived
class off of the \rate\ class to implement some of these approximations.  Very specialized networks like the {\tt iso7} network \citep{iso7} or rp-process approximate network of \citet{rprox} might require specialized code paths, but are still possible to implement to allow for interactive exploration and exporting the network in different formats with updated rates.

The python interface to nuclear reaction data and network structure allows for easier analysis of trends in reaction networks, like the work of \citet{Zhu2016,meyer:2018},  and the development of machine learning approaches to approximating nuclear kinetics \citep{fan:2022}.

We are also in the process of implementing two methods for reaction network reduction that were originally developed for the chemical reaction networks used in combustion simulations. The first method was originally outlined by \citet{sun:2010}, and the second was developed by \citet{pepiot-desjardins:2008} and explored further in \citet{niemeyer:2011}. The pyMARS library \citep{Mestas:2019} implements these methods for chemical networks, and inspired their inclusion in \pynucastro.

We also want to expand the range of nuclear physics data sources we encompass.  Since this is developed openly, on github, the hope is that others in the community will directly contribute new rates as they become available.  In the near term, we will increase our coverage of weak rates.  At the moment, we include those from \citet{suzuki2016}, since Urca was an initial application area.  The tabulated weak rates of \citet{fuller:1985}, \citet{langanke:2000}, and others
fit into the \tabularrate\ framework well.
This coverage will also allow us to generate NSE tables from large networks that include $Y_e$ evolution that can be used within hydro codes, as in \citet{ma:2013}, to capture the energetic and $Y_e$ evolution from a large network in tandem to a more reasonable sized network used for advection.  In this case, tabulating $dY_e/dt$ in addition to the NSE state would allow for a second-order accurate predictor-corrector time-integration scheme with NSE (especially when combined with the spectral deferred corrections integrator, as in \citealt{castro_simplified_sdc}).

Although our primary source for thermonuclear rates is \reaclib, other rate compilations
exist in the community, including \starlib~\citep{starlib}.  \starlib\ goes beyond \reaclib\ in including
error information associated with the tabulated rates.  Support for \starlib\ can be added in a straightforward manner by subclassing either \rate\ or \reaclibrate, and would be a good project for a new contributor.

It is also straightforward to increase the application codes
that we can output networks to.  We use a template find-and-replace strategy to insert the functions needed to
evaluate the network into the \cxx\ code functions that communicate with the \amrex-Astro simulation codes.  Similar templates can be used for other codes with few modifications to the code.  This would be a good project for a student using a simulation code in their research.

Finally, the focus of \pynucastro\ so far is on nuclear properties and networks, but support for more areas of nuclear astrophysics can be added as needed, such as wrappers for popular equations of state and neutrinos.  Plasma neutrino losses would make it easy to plot ignition curves, for example, and since we keep track of the individual rates, we can handle the neutrino losses from H-burning reactions in a straightforward manner.

\begin{acknowledgments}
We thank Jim Truran for teaching us all about nuclear astrophysics.
We thank Luna for her help sniffing out bugs in \pynucastro.  

The work at Stony Brook was supported by
DOE/Office of Nuclear Physics grant DE-FG02-87ER40317. This material is based
upon work supported by the U.S. Department of Energy, Office of Science, Office
of Advanced Scientific Computing Research and Office of Nuclear Physics,    
Scientific Discovery through Advanced Computing (SciDAC) program under Award
Number DE-SC0017955.  This research was supported by the Exascale Computing 
Project (17-SC-20-SC), a collaborative effort of the U.S. Department of Energy
Office of Science and the National Nuclear Security Administration. ASC is funded by the Chilean Government ANID grant number 56160017 and the U.S Department of State, Fulbright grant number PS00280789. 
\end{acknowledgments}

%


\software{AMReX \citep{amrex_joss},
          Jupyter \citep{jupyter},
          matplotlib \citep{Hunter:2007},
          NetworkX \citep{networkx},
          Numba \citep{numba},
          NumPy \citep{numpy,numpy2},
          pytest \citep{pytest},
          SymPy \citep{sympy}}


\appendix

\section{Ideal Boltzmann Gas Approximation}
In order to relate the different thermodynamics quantities computed in
a reaction network, an equation of state should be provided. In the
\pynucastro\ design we assumed the non-degenerate, non-interacting
and non-relativistic regime for the nuclei, with exceptions discussed
in section \ref{sec:screen}. As we increase the temperature, keeping
the density $\lesssim 10^{11}~\gcc$, the Boltzmann
statistics governs the behaviour of the nuclei involved in all the
reaction network. In this section we want to stress the relationship
between the number density $n$, chemical potential $\mu$ and
temperature $T$ under this approximation. These calculations are
standard and can be found in many references (see, for example,
\citealt{hill:1960}).

The grand-canonical partition function $\mathcal{Z}$ for $k$ subsystems is given by:
\begin{equation}\label{eq:can1}
    \mathcal{Z} = \sum_k (z^kZ_k)   
\end{equation}
where $Z_k$ is the canonical partition function for the subsystem with $k$ identical particles, $z=e^{\mu/ \kb T}$ is the fugacity parameter, and $\kb$ is the Boltzmann constant. From here:
\begin{equation}\label{eq:can2}
    Z_k = \dfrac{Z_{\mathrm{1\mbox{-}p}}^k}{k!}
\end{equation}
where $Z_{\mathrm{1\mbox{-}p}}$ is the canonical partition function for only one particle:
\begin{equation}\label{eq:can3}
    Z_{1\mbox{-}p} = \sum_l \int \mathrm{d}^3\mathbf{p}\; \mathrm{d}^3\mathbf{x} \;e^{- \epsilon(l,p_i,x_i)/\kb T}
\end{equation}
where $\mathbf{x}$ are the space parameters, $\mathbf{p}$ the momentum parameters, and $l$ labels the
internal degrees of freedom. For a spherically symmetric potential
bounded particle, the energy of each nuclei is given by
\begin{equation}\label{eq:can4}
    \epsilon(l, \mathbf{p},\mathbf{x}) = \dfrac{\mathbf{p}^2}{2m} + \Delta_l + mc^2 + \mu^C
\end{equation}
with $\mathbf{p}^2 = \mathbf{p} \cdot \mathbf{p}$ as the squared momentum magnitude, $\Delta_l$ as the
discrete energy levels of each particle, $mc^2$ as the rest-energy, and
$\mu^C$ as the Coulomb screening constant, which is the consequence of
the electron cloud interactions that surrounds once each nuclei is
ionized.

Combining the equations (\ref{eq:can1}--\ref{eq:can4}), we may compute $\mathcal{Z}$ by:
\begin{equation}
    \mathcal{Z} = \sum_{k} (2J_l + 1)e^{-\Delta_l/\kb T} V\left(\dfrac{m \kb T}{2\pi \hbar^2} \right)^{3/2} e^{(\mu-mc^2 - \mu^C)/\kb T} 
\end{equation}
where $V$ is the volume, $J_l$ is the $l$-energy level spin of the nuclei, and $m$ is the mass of the nuclei. Thus, the grand-canonical potential $\Omega$ is given by:
\begin{align}\label{ap_potential}
    \Omega &= - \kb T \log \mathcal{Z} \nonumber\\
    &= -\kb T \sum_l (2J_l + 1)e^{-\Delta_l/(\kb T)} V\left(\dfrac{m \kb T}{2\pi \hbar^2} \right)^{3/2} e^{(\mu - mc^2 - \mu^C)/\kb T} 
\end{align}
The number of particles as a function of $(T,V, \mu)$, where $\mu$ is the chemical potential, can be expressed as:
\begin{align} \label{eq:ap_particles}
    N &= - \left(\dfrac{\partial \Omega}{\partial \mu} \right)_{V,T} \nonumber\\
    &= V \left[  \sum_l (2J_l + 1)e^{-\Delta_l/\kb T}\right]\left(\dfrac{m \kb T}{2\pi \hbar^2} \right)^{3/2} e^{(\mu - mc^2 - \mu^C)/\kb T} 
\end{align}
After identifying the number density $n= N/V$, and introducing the normalized  partition function definition (\citealt{william:1967, rauscher:2003}):
\begin{equation}\label{eq:pf}
    (2J_0 + 1)G(T) = \sum_l (2J_l + 1)e^{-\Delta_l/\kb T}
\end{equation}
where $J_0$ is the nuclei ground-state spin, we can we can rewrite (\ref{eq:ap_particles}) in terms of $\mu$ as:
\begin{equation} \label{eq:ap_mu}
    \mu = \kb T \log \left[\dfrac{n}{(2J_0 + 1)G(T)} \left(\dfrac{2\pi \hbar^2}{m\kb T} \right)^{3/2} \right] + mc^2 + \mu^{C}
\end{equation}
Finally, we can define the kinematic term of the chemical potential by:
\begin{equation}\label{eq:ap_mu_kin}
    \mu^{\mathrm{id}} = \kb T \log \left[\dfrac{n}{(2J_0 + 1)G(T)} \left(\dfrac{2\pi \hbar^2}{m\kb T} \right)^{3/2} \right]
\end{equation}
suggesting $\mu = \mu^{\mathrm{id}} + mc^2 + \mu^{C}$. The case $\mu^{C}=0$ is the non-interacting or no-screening case. By computing the derivatives of (\ref{ap_potential}) it is possible to compute the entropy per ion and pressure in terms of $n$ and $T$ \citep{skynet}.

\section{Reverse Rates and Detailed Balance Calculations:}\label{sec:detailed_balance}

The relationship between the inverse and the forward reaction may be obtained by detailed balance calculations (see, for example, \citealt{rauscher:1997, reaclib, arnett, skynet, clayton, angulo:1999}).  Here we describe how the detailed balance calculations are performed. A nuclear reaction between a set of reactants $\{C_i\}$ and a set of products $\{B_j\}$ in equilibrium, is represented by:
\begin{equation}
    C_1 + C_2 + \cdots + C_r  \rightleftharpoons B_1 + B_2 + \cdots + B_p
\end{equation}
where $r$ and $p$ represent the number of reactants and products respectively. The right-arrow represents a forward reaction, while the left-arrow represents a reaction in the opposite direction known as the inverse reaction. After applying (\ref{eq:reaction_react}) to the forward reaction with $r$ reactants:
\begin{equation}\label{eq:ap_reverse_react}
     \dfrac{1}{c_{C_1}}\dfrac{dn_{C_1}}{dt} = \dfrac{1}{c_{C_2}}\dfrac{dn_{C_2}}{dt} = \cdots = \dfrac{1}{c_{C_r}}\dfrac{dn_{C_r}}{dt} = -\dfrac{n_{C_1}n_{C_2} \cdots n_{C_r}}{\prod_{C_i} c_{C_i}!} \langle\sigma v\rangle_{C_1 C_2 \cdots C_r}
\end{equation}
and to the inverse reaction with the $p$ products as reactants, implies that:
\begin{equation}\label{eq:ap_reverse_prod}
      \dfrac{1}{c_{B_1}}\dfrac{dn_{B_1}}{dt} = \dfrac{1}{c_{B_2}}\dfrac{dn_{B_2}}{dt} =  \cdots = \dfrac{1}{c_{B_p}}\dfrac{dn_{B_p}}{dt} = -\dfrac{n_{B_1}n_{B_2} \cdots n_{B_p}}{\prod_{B_i} c_{B_i}!} \langle\sigma v\rangle_{B_1 B_2 \cdots B_p}
\end{equation}
where $n_{C_i}$ is the nuclei $C_i$ number density, and $c_{C_i}$ is the count of the reactant nuclei $C_i$, and similarly, $c_{B_i}$ is the count of the product nuclei $B_i$. From (\ref{eq:ap_reverse_react}), (\ref{eq:ap_reverse_prod}) and the conservation of nucleons:
\begin{equation}\label{eq:ap_rev}
      \dfrac{ \langle\sigma v\rangle_{B_1 B_2 \cdots B_r}}{ \langle\sigma v\rangle_{C_1 C_2 \cdots C_r}} = \dfrac{n_{C_1}n_{C_2} \cdots n_{C_r}}{n_{B_1}n_{B_2} \cdots n_{B_p} }\times \dfrac{\prod_{B_i} c_{B_i}!}{\prod_{C_i} c_{C_i}!}
\end{equation}
Furthermore, after computing $n_{N_i}$ in terms of $\mu_{N_i}$, from (\ref{eq:ap_mu}), for each $N_i$-th nuclei, and using the approximation $m_{N_i} = A_{N_i}m_u$:
\begin{equation}\label{eq:ap_n_dens}
    n_{N_i} = (2J_0 + 1)A_{N_i}^{3/2}\left(\dfrac{m_u k_B T}{2\pi \hbar^2} \right)^{3/2} e^{(\mu_{N_i} - m_{N_i}c^2 - \mu^C_{N_i})/k_B T} G_{N_i}(T) .
\end{equation}
From (\ref{eq:ap_rev}) and (\ref{eq:ap_n_dens}) we obtain, depending on the value of $\mu^C$, a relationship between the forward and reverse rate. The case $\mu^C=0$ is the no-screening case, where $\mu^C \neq 0$ implies the presence of coulomb screening. \\

\textbf{Case $\mu^C = 0$:}
In this case, the previous substitution leads to:
\begin{align}
     \dfrac{N_a^{p-1}\langle\sigma v\rangle_{B_1 B_2 \cdots B_r}}{N_a^{r-1} \langle\sigma v\rangle_{C_1 C_2 \cdots C_r}} &=\left(\frac{1}{N_a}\right)^{r-p}  \dfrac{(2J_{C_1} + 1)\cdots (2J_{C_r} + 1)}{(2J_{B_1} + 1)\cdots (2J_{B_p} + 1)} \left(\frac{A_{C_1}\cdots A_{C_r} }{A_{B_1}\cdots A_{B_p} } \right)^{3/2} \left(\frac{m_u\kb}{2\pi \hbar^2} \right)^{\frac{3}{2}(r-p)}T^{\frac{3}{2}(r-p)} \nonumber\\
     &\hspace{0.5cm} \times \, \frac{G_{C_1} \cdots G_{C_r}(T)}{G_{B_1} \cdots G_{B_p}(T)} \dfrac{\prod_{B_i} c_{B_i}!}{\prod_{C_i} c_{C_i}!}\exp \left[-\dfrac{1}{\kb T}\left(\sum_{i=1}^r m_{C_i} - \sum_{i=1}^p m_{B_i}\right)c^2 \right]  \nonumber\\
     &\hspace{0.5cm}\times \,\exp \left[\dfrac{1}{\kb T}\left(\sum_{i=1}^r \mu_{C_i} - \sum_{i=1}^p \mu_{B_i} \right) \right]
\end{align}
where $N_a$ is the Avogadro's number, $J_{N_i}$ is the spin-value of the $N_i$-th nuclei ground state, $c_{N_i}$ is the count of nuclei $N_i$ in the reaction and $G_{N_i}(T)$ is the ground spin-state normalized nuclear partition function (\ref{eq:pf}), with $\Delta_{N_i,l}$ and $J_{N_i,l}$ as the energy-spectrum and the $l$-energy level spin of the nuclei $N_i$ respectively. The quantities $G_{N_i}$ are constructed and tabulated between $10^7\,\mathrm{K}$ and $10^{11}\,\mathrm{K}$ in \citet{rauscher:1997, rauscher:2003}. In \pynucastro\ we import these tables and convert them in a suitable format to be merged, consequently, no calculations of (\ref{eq:pf}) are performed. Each $A_{n_i}$ represents the atomic weight of the nuclei $n_i$, and we define $Q$ as the capture energy of the reaction, by:
\begin{equation}\label{eq:Q}
    Q = \left(\sum_{i=1}^r m_i - \sum_{i=1}^p m_p \right)c^2 .
\end{equation}
Since the reaction is in equilibrium, we can assert that:
\begin{equation}\label{eq:det_bal_mu}
    \sum_{i=1}^r \mu_{C_i} - \sum_{i=1}^p \mu_{B_i} = 0
\end{equation}
Using (\ref{eq:Q}) and (\ref{eq:det_bal_mu}) we can write the detailed balance inverse rate of the forward rate $N_a^{r-1}\langle \sigma v\rangle_{C_1 C_2 \cdots C_r}$ by: 
\begin{align}\label{eq:reverse}
N_a^{p-1}\left\langle \sigma v\right\rangle_{B_1,B_2,\cdots,B_p} &= \left(\frac{1}{N_a}\right)^{r-p} \frac{(2J_{C_1}+1) \cdots (2J_{C_r} + 1)}{ (2J_{B_1} + 1) \cdots (2J_{B_p} + 1)}\left(\frac{A_{C_1}\cdots A_{C_r} }{A_{B_1}\cdots A_{B_p} } \right)^{3/2}  \left(\frac{m_u\kb}{2\pi \hbar^2} \right)^{\frac{3}{2}(r-p)} T^{\frac{3}{2}(r-p)} \nonumber\\
 &\hspace{0.5cm}\times\, \frac{G_{C_1} \cdots G_{C_r}(T)}{G_{B_1} \cdots G_{B_p}(T)} \dfrac{\prod_{B_i} c_{B_i}!}{\prod_{C_i} c_{C_i}!}\mathrm{e}^{-{Q}/{\kb T}} N_a^{r-1}\left\langle \sigma v\right\rangle_{C_1,C_2,\cdots,C_r},
\end{align}
From (\ref{eq:reverse}), the determination of the quantities $J_{n_i}$ and $Q$ becomes essential in the goal of computing the reverse rates. In \reaclib, each rate $N_a^{n-1}\langle \sigma v \rangle_{n_1,n_2\cdots,n_n}$ is represented by seven parameters described in (\ref{eq:reaclib_fit}). After setting each partition function by $G_{n_i}(T)=1$ in equation (\ref{eq:reverse}), replacing
(\ref{eq:reaclib_fit}) and taking the logarithm on both sides, we may rewrite (\ref{eq:reverse}) as (\ref{eq:det_balance}), where $N_a^{p-1}\langle \sigma v \rangle_{B_1,B_2\cdots,B_p}'$ should satisfy the fit (\ref{eq:reaclib_fit}) with the set $a_{0,\mathrm{rev}}, \cdots, a_{6,\mathrm{rev}}$ of reverse coefficients described in \citet{rauscher:1997}, but generalized for $r$ reactants and $p$ products as:
\begin{subequations}
\begin{align}
    a_{0, \mathrm{rev}} &= \log \left \{\frac{(2J_{C_1}+1) \cdots (2J_{C_r} + 1)}{ (2J_{B_1} + 1) \cdots (2J_{B_p} + 1)}\left(\frac{A_{C_1}\cdots A_{C_r} }{A_{B_1}\cdots A_{B_p} } \right)^{3/2} \dfrac{\prod_{B_i} c_{B_i}!}{\prod_{C_i} c_{C_i}!} F \right\} \,+ a_0 \label{eq:a0_rev}\\
    a_{1,\mathrm{rev}} &= a_1 - Q/(10^9\,\mathrm{K}\cdot\kb)\\
    a_{2,\mathrm{rev}} &= a_2\\
    a_{3,\mathrm{rev}} &= a_3\\
    a_{4,\mathrm{rev}} &= a_4\\
    a_{5,\mathrm{rev}} &= a_5\\
    a_{6,\mathrm{rev}} &= a_6 + \frac{3}{2}(r-p) \label{eq:a6_rev}
\end{align}
\end{subequations}
with $F$ defined by:
\begin{equation}
    T_9^{\frac{3}{2}(r-p)}F = \left(\dfrac{1}{N_a} \right)^{r-p}\left( \dfrac{m_u \kb}{2\pi \hbar^2} \right)^{\frac{3}{2}(r-p)} T^{\frac{3}{2}(r-p)}
\end{equation}
\textbf{Case $\mu^C \neq 0$:}
In this case, the rates are screened and due to $\mu^C \neq 0$, we have to incorporate the screening interactions in the detailed balance calculations. We may find an example of these calculations in \citealt{SEITENZAHL200996, Calder_2007}. From (\ref{eq:reverse}) and (\ref{eq:ap_n_dens}):
\begin{align} 
    N_a^{p-1}\left\langle \sigma v\right\rangle_{B_1,B_2,\cdots,B_p}^{\mathrm{sc}} &= \left(\frac{1}{N_a}\right)^{r-p} \frac{(2J_{C_1}+1) \cdots (2J_{C_r} + 1)}{ (2J_{B_1} + 1) \cdots (2J_{B_p} + 1)}\left(\frac{A_{C_1}\cdots A_{C_r} }{A_{B_1}\cdots A_{B_p} } \right)^{3/2}  \left(\frac{m_u\kb}{2\pi \hbar^2} \right)^{\frac{3}{2}(r-p)} T^{\frac{3}{2}(r-p)} \nonumber\\
 &\hspace{0.5cm}\times\, \frac{G_{C_1} \cdots G_{C_r}(T)}{G_{B_1} \cdots G_{B_p}(T)} \dfrac{\prod_{B_i} c_{B_i}!}{\prod_{C_i} c_{C_i}!}\mathrm{e}^{-{Q}/{\kb T}} \nonumber \\
 &\hspace{0.5cm}\times\, \exp \left[-\dfrac{1}{\kb T}\left(\sum_{i=1}^r \mu_{C_i}^C - \sum_{i=1}^p \mu_{B_i}^C \right) \right] N_a^{r-1}\left\langle \sigma v\right\rangle_{C_1,C_2,\cdots,C_r}^{\mathrm{sc}} \label{eq:reverse_sc}
\end{align}
This equation will become essential for the next section discussion.

\section{Electron Screening: Theoretical Considerations}

The Coulomb screening enhancement factor (\ref{eq:f_factor}) of a reaction can be decomposed as the product of the enhancement factors of the intermediate reactions. Similarly, from (\ref{eq:reverse_sc}) we can assert that the enhancement factor of an inverse detailed balance reactions replaces the reactants by the products in (\ref{eq:f_factor}). Let us explore these interesting features of the factor (\ref{eq:f_factor}). From the following chain of rates:
\begin{equation}
    C_1 + C_2 + \cdots + C_r  \rightarrow  [C_1 + C_2 + \cdots + C_{r-1} ] + C_r \rightarrow  \  B_1 + B_2 + \cdots + B_p
\end{equation}
where the brackets $[\cdots]$ denotes a composite nuclei obtained as a result of the $r-1$ original $C_1,C_2, \cdots C_{r-1}$ nuclei collisions, we have:
\begin{align}\label{eq:sc_map}
    f_{C_1 \cdots C_{r-1}}f_{[C_1 + \cdots +C_{r-1}], C_r} & = e^{[\mu^C(Z_{C_1}) + \cdots + \mu^C(Z_{C_{r-1}}) - \mu^C(Z_{C_1} + \cdots Z_{C_{r-1}})]/\kb T} \nonumber\\
    &\hspace{1cm}\times e^{[\mu^C(Z_{C_1} + \cdots + Z_{C_{r-1}}) + \mu^C(Z_{C_r}) - \mu^C(Z_{C_1} + \cdots Z_{C_r})]/\kb T} \nonumber \\
    &=e^{[\mu^C(Z_{C_1}) + \cdots + \mu^C(Z_{C_r}) - \mu^C(Z_{C_1} + \cdots Z_{C_r})]/\kb T} \nonumber\\
    &= f_{C_1 \cdots C_r}
\end{align}
Therefore, independent of the number of reactants, each rate can be screened in sequences of pairs until all the remaining reactants are exhausted. Using this property, screening factors of reactions like the 3-$\alpha$ chain:
\begin{equation}
    \alpha + \alpha + \alpha \rightarrow \isotm{Be}{8} + \alpha \rightarrow \isotm{C}{12}
\end{equation}
may be computed in two steps: first, we compute $f_{\alpha \alpha}$ and $f_{\isotm{Be}{8}, \alpha}$; and then multiply them to obtain the original rate $f_{\alpha \alpha \alpha}$ screening factor.

The quantities $\mu^C_i$ required to compute the screening factors are also required to perform the NSE state calculations of the network, as shown in (\ref{eq:ap_nucleon_fraction}). Therefore the electron screening not only enhance the reaction network convergence, but also modify its NSE state. Similarly, according to the extra term $\mu^C$ from (\ref{eq:ap_mu}), the reverse reactions are also modified by the screening effects into (\ref{eq:reverse_sc}), altering the performance of the network convergence. Let us examine how the detailed balance calculations (\ref{eq:reverse}) are modified by the inclusion of $\mu^C$ in (\ref{eq:reverse_sc}), where 
\begin{equation}
N_a^{r-1}\left\langle \sigma v\right\rangle_{C_1,C_2,\cdots,C_r} ^{\mathrm{sc}}=N_a^{r-1}\left\langle \sigma v\right\rangle_{C_1,C_2,\cdots,C_r} ^{\mathrm{no-sc}}\times \mathrm{e}^{\left(\sum_{C_i} \mu_{C_i}^C - \mu^C_{C_1 + C_2 + \cdots +C_r} \right)/\kb T} 
\end{equation}
is the forward screened rate expressed in terms of the forward unscreened rate by (\ref{eq:f_factor}). Substituting this into (\ref{eq:reverse_sc}):
\begin{align}\label{eq:reverse_sc_1}
    N_a^{p-1}\left\langle \sigma v\right\rangle_{B_1,B_2,\cdots,B_p}^{\mathrm{sc}} &= \left(\frac{1}{N_a}\right)^{r-p} \frac{(2J_{C_1}+1) \cdots (2J_{C_r} + 1)}{ (2J_{B_1} + 1) \cdots (2J_{B_p} + 1)}\left(\frac{A_{C_1}\cdots A_{C_r} }{A_{B_1}\cdots A_{B_p} } \right)^{3/2}  \left(\frac{m_u\kb}{2\pi \hbar^2} \right)^{\frac{3}{2}(r-p)} T^{\frac{3}{2}(r-p)} \nonumber\\
 &\hspace{0.5cm}\times\, \frac{G_{C_1} \cdots G_{C_r}(T)}{G_{B_1} \cdots G_{B_p}(T)} \dfrac{\prod_{B_i} c_{B_i}!}{\prod_{C_i} c_{C_i}!}\mathrm{e}^{-{Q}/{\kb T}} N_a^{r-1}\left\langle \sigma v\right\rangle_{C_1,C_2,\cdots,C_r} ^{\mathrm{no-sc}} \nonumber\\
 &\hspace{0.5cm}\times \mathrm{e}^{\left(\sum_{C_i} \mu_{C_i}^C - \mu^C_{C_1 + C_2 + \cdots +C_r} \right)/\kb T} \mathrm{e}^{\left(-\sum_{C_i} \mu_{C_i}^C + \sum_{B_i} \mu_{B_i}^C \right)/\kb T}
\end{align}
and from the conservation of charge $Z_{C_1} + \cdots + Z_{C_r} = Z_{B_1} + \cdots + Z_{B_p}$:
\begin{equation}\label{eq:reverse_sc_2}
    \mu_{C_1 + \cdots + C_r}^C = \mu_{B_1 + \cdots + B_p}^C 
\end{equation}
we can write, from (\ref{eq:reverse_sc_1}), (\ref{eq:reverse_sc_2}) and by identifying $ N_a^{p-1}\left\langle \sigma v\right\rangle_{B_1,B_2,\cdots,B_p}^{\mathrm{no-sc}}$ from (\ref{eq:reverse}):
\begin{align}
    N_a^{p-1}\left\langle \sigma v\right\rangle_{B_1,B_2,\cdots,B_p}^{\mathrm{sc}} &= N_a^{p-1}\left\langle \sigma v\right\rangle_{B_1,B_2,\cdots,B_p}^{\mathrm{no-sc}} \times \mathrm{e}^{\left(\sum_{C_i} \mu_{C_i}^C - \mu^C_{C_1 + C_2 + \cdots + C_r} \right)/\kb T} \mathrm{e}^{\left(-\sum_{C_i} \mu_{C_i}^C + \sum_{B_i} \mu_{B_i}^C \right)/\kb T} \nonumber\\
    &= N_a^{p-1}\left\langle \sigma v\right\rangle_{B_1,B_2,\cdots,B_p}^{\mathrm{no-sc}} \times \mathrm{e}^{\left(\sum_{B_i} \mu_{B_i}^C - \mu^C_{B_1 + B_2 + \cdots + B_r} \right)/\kb T} \nonumber \\
    &= N_a^{p-1}\left\langle \sigma v\right\rangle_{B_1,B_2,\cdots,B_p}^{\mathrm{no-sc}}\times f_{B_1 \cdots B_p} \label{eq:f_factor_inv}
\end{align}
where $f_{B_1 \cdots B_p}$ is the screening factor of the reverse reaction rate. This is a very powerful result, because each rate, independent of its forward/inverse nature, may be computed by using its reactants/products respectively. Therefore, in order to implement screening in detailed balanced reverse rates we need to just consider the reversed rate in the definition of the enhancement factor (\ref{eq:f_factor}), leaving the unscreened detailed balance calculations unchanged.

\section{Computing the NSE State} \label{sec:nse_derivation}

The nuclear statistical equilibrium (NSE) state of a network, describes a unique composition in which each nucleus's set of protons and neutrons are in equilibrium if the protons and neutrons that assemble the nuclei were free, given a set of thermodynamic state variables and electron fraction. (see an alternative derivation in \citealt{Kushinr}).
In this state, the energy required to assemble the $i$-th nuclei is just the energy required to setup $Z_i$ free protons and $N_i$ free neutrons \citep{Clifford-and-tayler}. This condition is given by:
\begin{equation}
    \mu_i = Z_i \mu_p + N_i \mu_n
\end{equation}
Using (\ref{eq:ap_mu}):
\begin{equation}
    \mu_i^{\mathrm{id}} + m_ic^2 + \mu_i^C = Z_i (\mu_p^{\mathrm{id}} + m_pc^2 + \mu_p^C ) + N_i(\mu_n^{\mathrm{id}} + m_nc^2 + \mu_n^C )
\end{equation}
or equivalently, after setting $\mu^C_n$ = 0, because neutrons are not charged, and introducing the binding energy of the $i$-th nuclei:
\begin{equation}
    Q_i=(Z_im_p + N_im_n - m_i)c^2
\end{equation}
we can write:
\begin{equation} \label{eq:ap_nse1}
    \mu_i^{\mathrm{id}} = Z_i \mu_p^{\mathrm{id}} + N_i\mu_n^{\mathrm{id}} + Q_i - \mu_i^C + Z_i\mu^C_p
\end{equation}
From (\ref{eq:ap_nse1}), (\ref{eq:ap_mu_kin}), and using the nucleon fraction $X_i$ definition $n_i=\rho X_i / m_i$, we may finally compute (\ref{eq:ap_nucleon_fraction}).

\section{The \lowercase{\subchapprox} network}
\label{app:subch_approx}

The \subchapprox\ network is designed to model explosive helium
and carbon burning, containing the nuclei in the standard {\tt aprox13} network, as well as other nuclei and pathways identified in \citet{shenbildsten} for bypassing the $\isotm{C}{12}(\alpha,\gamma)\isotm{O}{16}$ rate. The original
version of this network appeared in \citet{castro_simplified_sdc}.  This approximate version approximates some of the $(\alpha,p)(p,\gamma)$ rates, as describing in section \ref{sec:approximate}.  It also modifies rates, as described in that
same section, to change the endpoints, assuming fast neutron captures.  It is generated as follows:

\begin{lstlisting}
rl = pyna.ReacLibLibrary()

nucs = ["p", "he4",
        "c12", "o16", "ne20",
        "mg24", "si28", "s32",
        "ar36", "ca40", "ti44",
        "cr48", "fe52", "ni56",
        "al27", "p31", "cl35",
        "k39", "sc43", "v47",
        "mn51", "co55",
        "n13", "n14", "f18",
        "ne21", "na22", "na23"]

subch = rl.linking_nuclei(nucs)

other = [(("c12", "c12"), ("mg23", "n"), ("mg24")),
         (("o16", "o16"), ("s31", "n"), ("s32")),
         (("c12", "o16"), ("si27", "n"), ("si28"))]

for r, p, mp in other:
    rfilter = pyna.RateFilter(reactants=r,
                              products=p)
    _library = rl.filter(rfilter)
    r = _library.get_rates()[0]
    r.modify_products(mp)
    subch += _library

net = pyna.PythonNetwork(libraries=[subch],
                         symmetric_screening=True)
net.make_ap_pg_approx(intermediate_nuclei=["cl35", "k39", "sc43",
                                           "v47", "mn51", "co55"])
net.remove_nuclei(["cl35", "k39", "sc43",
                   "v47", "mn51", "co55"])
\end{lstlisting}

\begin{figure*}[t]
\plotone{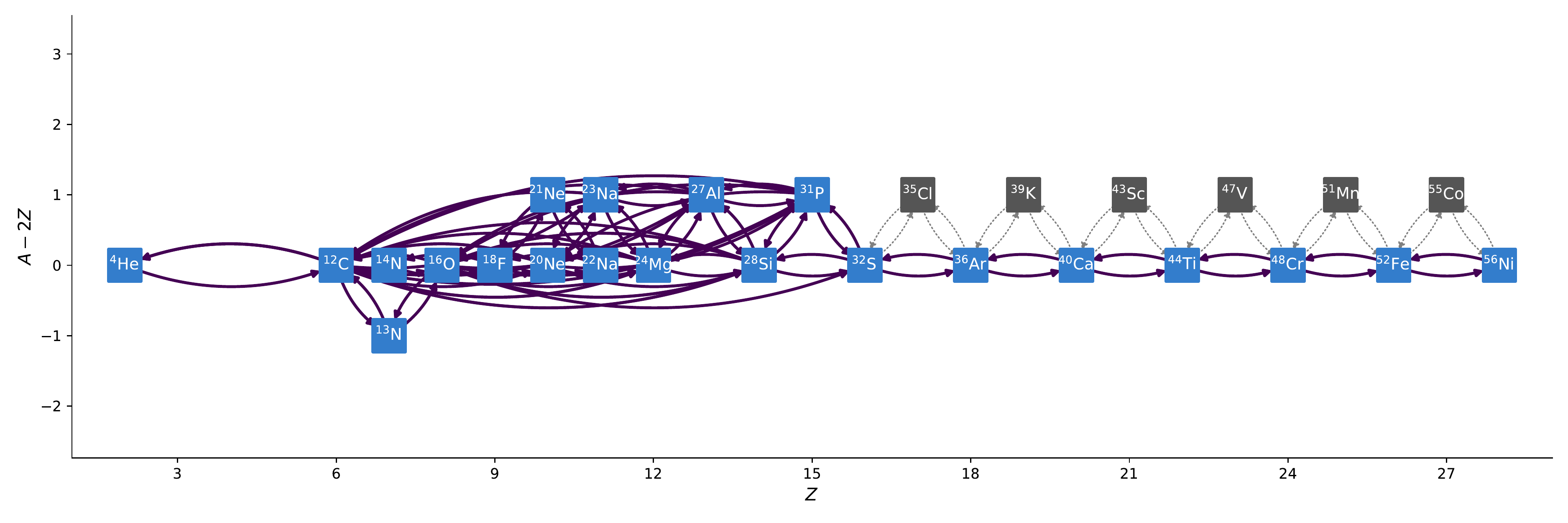}
\caption{\label{fig:subch_approx} The \subchapprox\ network.}
\end{figure*}

This network is visualized in Figure~\ref{fig:subch_approx}.

\end{document}